\documentclass{sig-alternate}

\usepackage{graphicx}   
\usepackage{balance}    
\usepackage{subfigure}
\usepackage{multirow}
\usepackage{pbox}
\usepackage{cite}
\usepackage{float}
\usepackage[table]{xcolor}
\usepackage{amsmath}
\usepackage{hhline}
\usepackage[font={small}]{caption}
\newcommand\abs[1]{\left|#1\right|}

\newcommand{\figscale}{0.9}

\makeatletter

\begin{document}

\title{Dejavu: An Accurate Energy-Efficient Outdoor Localization System}

\numberofauthors{2}
\author{
\alignauthor Heba Aly\\
  	\affaddr{Dept. of Computer and Systems Engineering}\\
  	\affaddr{Alexandria University, Egypt}\\
  	\email{heba.aly@alexu.edu.eg}
\alignauthor Moustafa Youssef\\
  	\affaddr{Wireless Research Center}\\
  	\affaddr{Alexandria University  and E-JUST, Egypt}\\
  	\email{moustafa.youssef@ejust.edu.eg}
}

\maketitle
\begin{abstract}
We present \emph{Dejavu}, a system that uses standard cell-phone sensors to provide accurate and energy-efficient outdoor localization suitable for car navigation. Our analysis shows that different road landmarks have a unique signature on cell-phone sensors; For example, going inside tunnels, moving over bumps, going up a bridge, and even potholes all affect the inertial sensors on the phone in a unique pattern. \emph{Dejavu} employs a dead-reckoning localization approach and leverages these road landmarks, among other automatically discovered abundant virtual landmarks, to reset the accumulated error and achieve accurate localization. To maintain a low energy profile, \emph{Dejavu} uses only energy-efficient sensors or sensors that are already running for other purposes.

We present the design of \emph{Dejavu} and how it leverages crowd-sourcing to automatically learn virtual landmarks and their locations. Our evaluation results from implementation on different android devices in both city and highway driving show that \emph{Dejavu} can localize cell phones to within 8.4~m median error in city roads and 16.6~m on highways. Moreover, compared to GPS and other state-of-the-art systems, \emph{Dejavu} can extend the battery lifetime by 347\%, achieving even better localization results than GPS in the more challenging in-city driving conditions.
\end{abstract}



\begin{figure*}[!ht]
\centering
\includegraphics[width=0.75\linewidth]{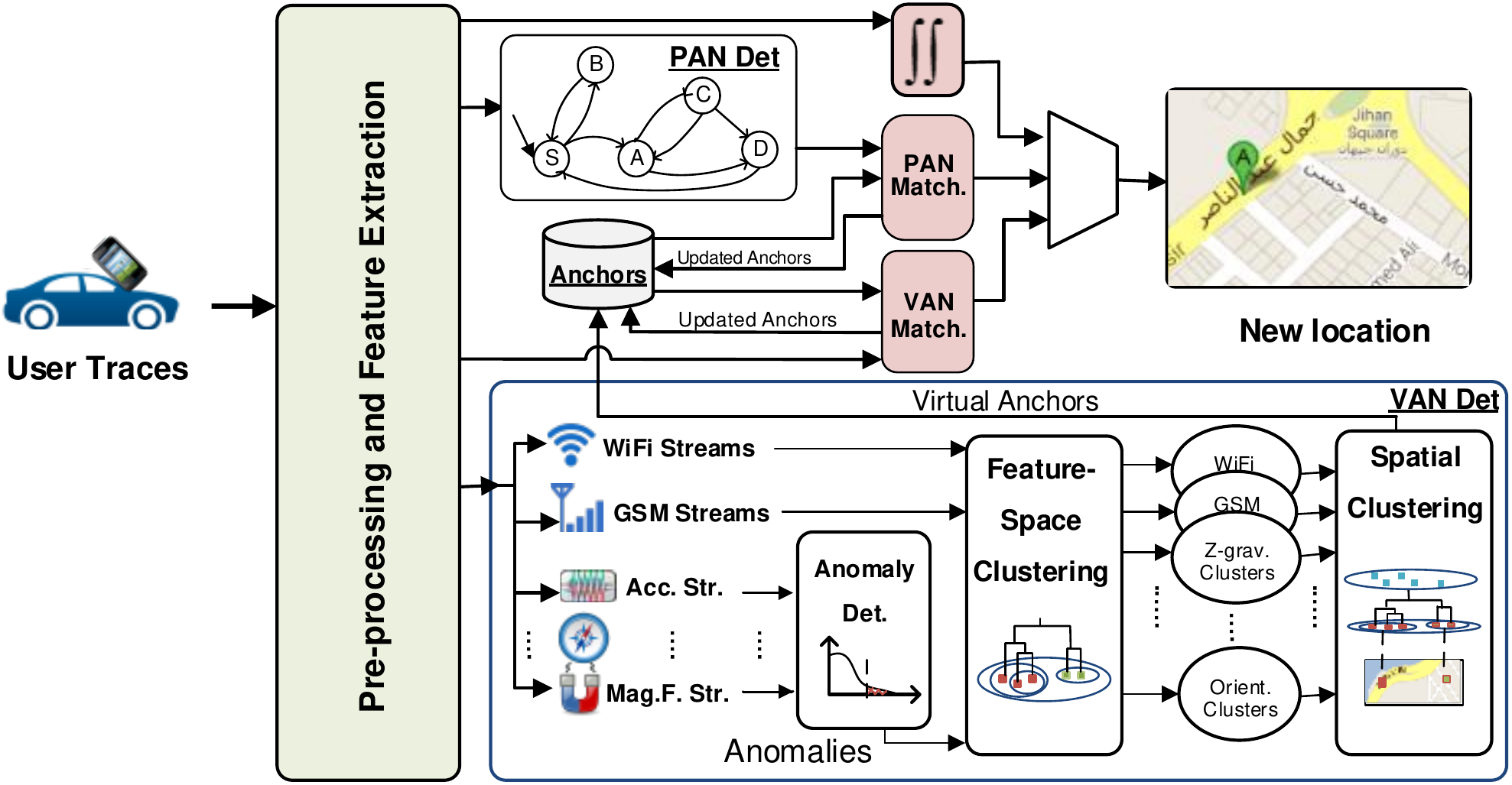}
\caption{\emph{Dejavu} architecture --- The phone location is estimated using dead-reckoning, physical (PAN) and virtual (VAN) anchors are used to reset accumulated error. Sensor traces are also mined to detect new anchors, improving the system accuracy over time.}
\label{fig:arch}
\end{figure*}

\section{Introduction}\label{sec:intro}
Location-based services (LBS) have become an integral part of our daily life with applications including car navigation, location-based social networks, and context-aware predication and advertisement. Different LBS require different localization accuracies; Generally, GPS is considered the de facto standard for ubiquitous and accurate outdoor navigation. However, GPS is an energy-hungry technology that can drain the scarce battery resource of mobile devices quickly. In addition, its accuracy is limited in areas with obscured access to the satellites, e.g. in tunnels and many urban areas.

To address the high energy requirement of GPS-based localization, a number of outdoor localization systems have been proposed over the years \cite{GAC, CompAcc, skyhook, enloc}. %
   For example, city-wide WiFi and cellular-based localization systems depend on fingerprinting the WiFi and cellular  networks through a war-driving process to remove the need for GPS. Other systems, e.g. \cite{GAC,CompAcc}, depend on the inertial sensors in today's smart phones to obtain the location. through a dead-reckoning approach and revert to GPS sampling with a low duty cycle to reset the accumulated localization error. However, this saving in energy usually comes at a reduced localization accuracy, affecting the range of possible LBS.

In this paper, we present \emph{Dejavu} as a system capable of providing both \textbf{accurate and energy-efficient} outdoor localization. At the core of \emph{Dejavu}, we use a dead-reckoning approach based on the low-energy profile inertial sensors (i.e. the accelerometer, compass, and gyroscope). However, using the array of sensors available in today's cell phones, \emph{Dejavu} identifies unique points in the environment, i.e. landmarks or anchors, and uses them to reset the error accumulation in the dead-reckoning displacement; Bridges, tunnels, curves, and even potholes all have unique sensor signatures and represent frequent error-resetting opportunities along the road. For example, when the car goes over a bump, \emph{Dejavu} detects the bump signature on the sensors and resets the car location to the bump location. \emph{Dejavu} constructs a database of multi-model sensor anchors 
 and leverages it to achieve accurate outdoor \textbf{\emph{car}} localization. To maintain energy-efficiency, \emph{Dejavu} depends on energy-efficient sensors as well as sensors that are already running for other purposes, e.g. GSM and opportunistic WiFi signal strength.

To build the anchor database, \emph{Dejavu} uses a crowd-sourcing approach, where cell phones contribute their sensor information and location estimate. \emph{Dejavu} analyzes these sensor readings to detect physical anchors, such as bridges and tunnels, as well as virtual anchors (e.g. points with a unique cellular signal strength signature). 

 Implementation of \emph{Dejavu} over android phones shows that it can provide outdoor car localization with a median accuracy of 8.4m inside cities and 16.6m in highways. A phone running \emph{Dejavu} drains power
347\% more efficiently than GPS. In addition, \emph{Dejavu} can provide even better accuracy than GPS inside cities.

In summary, our contributions are summarized as follows:
\begin{itemize}
\item We present the architecture of \emph{Dejavu}: a system that combines dead-reckoning with sensed road anchors to provide accurate and energy-efficient outdoor localization suitable for car navigation.

\item We provide a framework for detecting unique outdoor landmarks in the environment based on the phone sensors. A unified finite state machine approach is used to detect bootstrapping physical anchors (e.g. tunnels, bridges, turns, bumps, etc). In addition, unsupervised learning techniques are used to further detect virtual anchors (landmark with unique sensors signatures) and to automatically and transparently grow  the landmark database in a crowd-sourcing approach.
\item We implement our system on android-based phones and evaluate its performance as compared to the state-of-the-art systems, in different scenarios.
\end{itemize}

The rest of the paper is organized as follows: Section~\ref{sec:sysov} presents the system architecture. Section~\ref{sec:designdet} gives the details of the \emph{Dejavu} system. We provide the implementation and evaluation of \emph{Dejavu} in Section~\ref{sec:eval}. Section~\ref{sec:relwork} discusses related work. Finally, Section~\ref{sec:conclude} provides concluding remarks.

\section{System Overview}\label{sec:sysov}

Figure~\ref{fig:arch} shows the system architecture. \emph{Dejavu} has two main modules: anchor detection and location fusion. The system estimates the location of a cell phone that is attached to the car windshield or dashboard using a dead-reckoning approach, where the new location is calculated based on the previous location and displacement (calculated from the accelerometer) along the direction of motion (estimated from the compass and gyroscope). Due to the noisy inertial sensors, error accumulates with time. To reduce this error accumulation, \emph{Dejavu} automatically builds and leverages an anchor database, based on the collected sensors information, to reset the error; Whenever the phone detects an anchor signature, the phone location is adjusted to the anchor location and error is reset (Figure~\ref{fig:error_int}).

\subsection{Raw Sensor Information}
The system collects raw sensor information from cell phones. These include inertial sensors\footnote{The inertial sensor data is a vector quantity along the three dimensions. We use the individual components as well as the combined magnitude as input data.} as well as cellular network information (associated cell tower ID and its received signal strength (RSS) plus neighbouring cell towers and associated RSS). These sensors have the advantage of having a low cost energy profile and being run all the time during the phone operation, to maintain cellular connectivity or to detect phone orientation changes. Therefore, using them for localization consumes zero extra energy. In addition, \emph{Dejavu} opportunistically leverages the WiFi chip, if enabled, to collect surrounding WiFi APs. Note that although we focus on low-energy profile sensors in this paper, extending \emph{Dejavu} to use other sensors (such as the mic, camera, temperature, etc) is straightforward.

\subsection{Dead-reckoning and Anchor-based Error Resetting}
\emph{Dejavu} uses the linear acceleration combined with the direction of motion $\theta$ to compute the phone displacement based on its previous location. The acceleration signal is double integrated to obtain the displacement and the compass is used to obtain the direction of motion ($\theta$).
To compute the new phone position, we use the Vincenty's formula, rather than the Euclidian distance, as it takes the Earth curvature into account~\cite{GAC}.

\begin{figure}[!t]
\centering
\includegraphics[width=\figscale\linewidth]{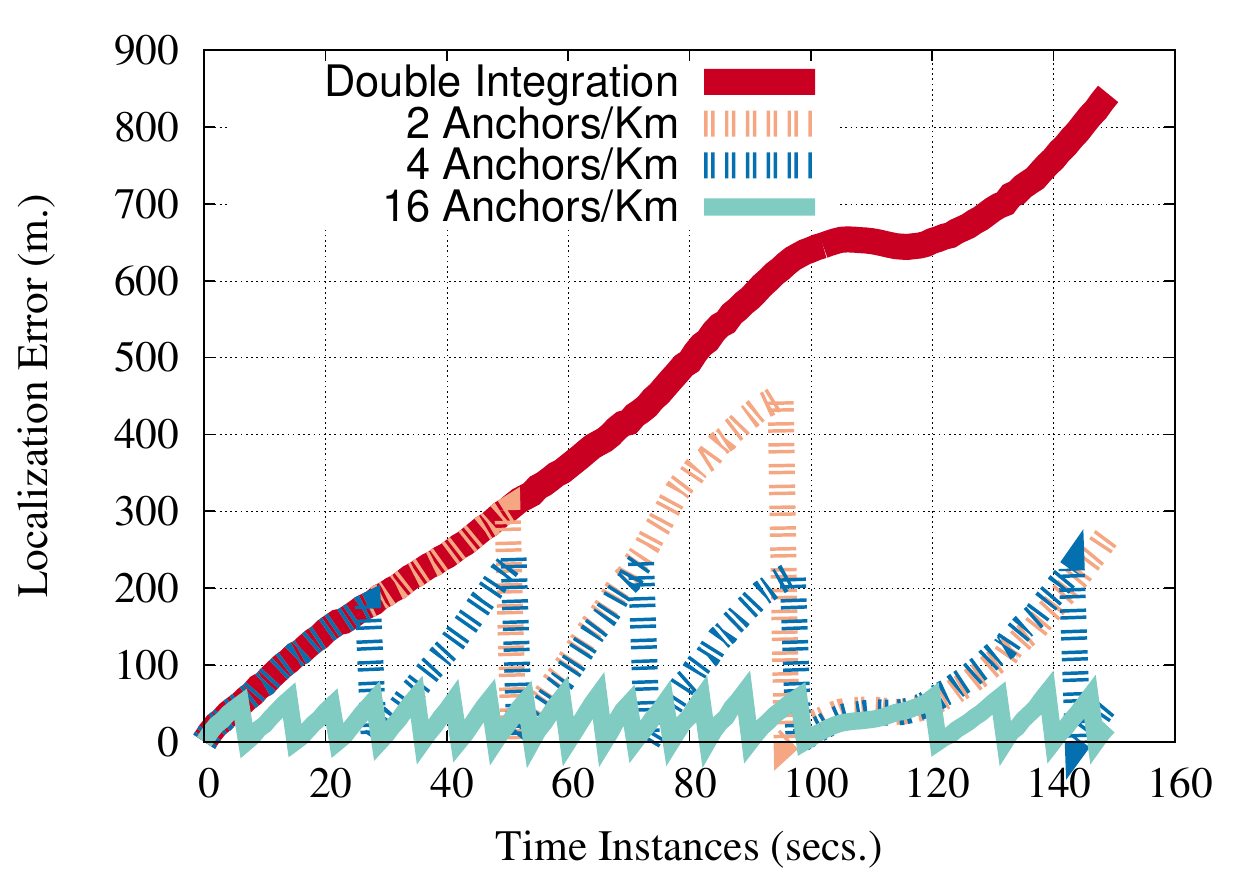}
\caption{Comparing double integration (dead-reckoning) error to \emph{Dejavu} with different anchor densities per kilo meter---The trip length is 1.5 Km.}
\label{fig:error_int}
\end{figure}

Due to the noise in the accelerometer and compass readings, error accumulates as time goes on. To limit the accumulated error, \emph{Dejavu} uses both physical and virtual anchors along the road to reset the localization error (Figure~\ref{fig:error_int}). The higher the density of anchors, the more frequent the resetting opportunities, and hence the higher the accuracy.

\subsection{Anchors Detection}
\begin{figure}[!t]
\centering
\includegraphics[width=\figscale\linewidth]{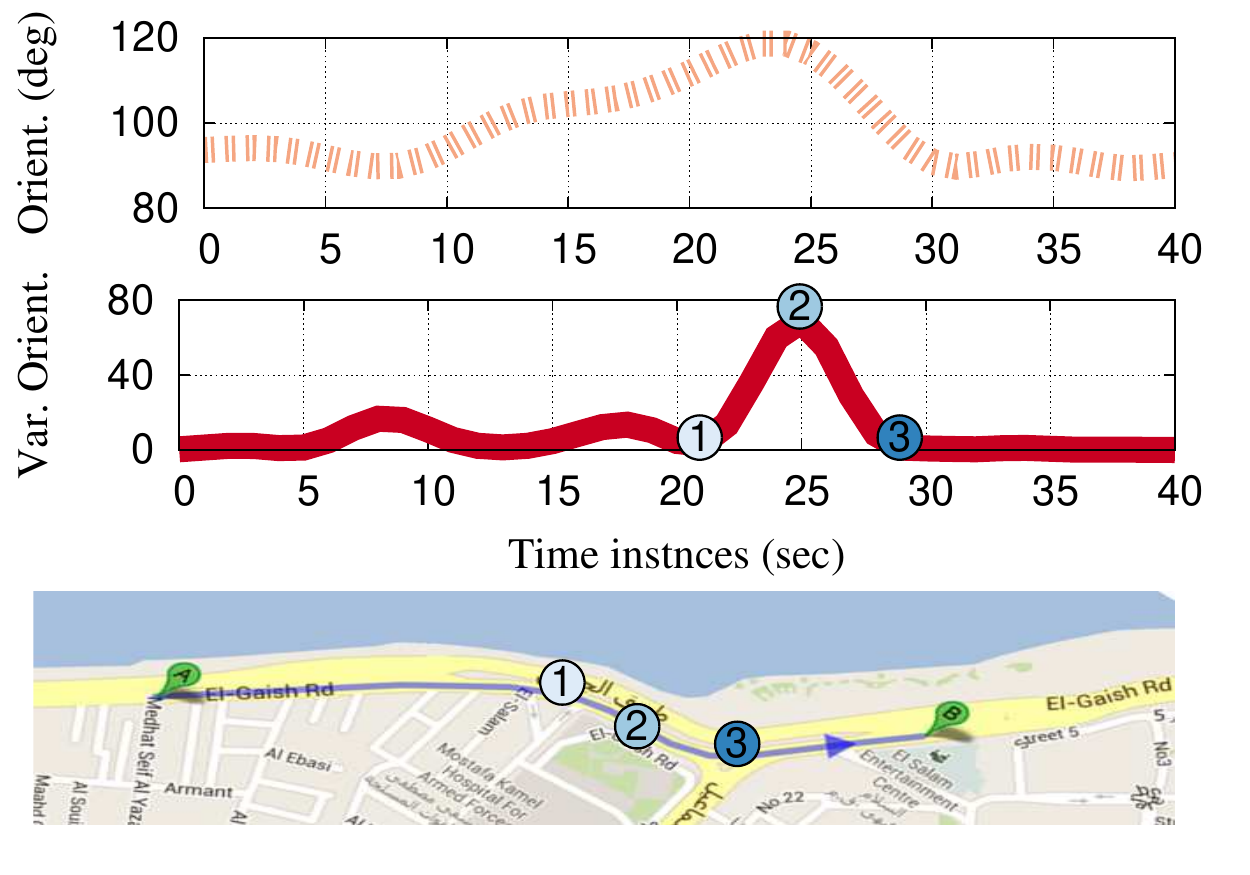}
\caption{A car moving along a curve affects the orientation angle. The variance of the orientation can be used to detect different numbered landmarks.}
\label{fig:slm_ex1}
\end{figure}

Our analysis showed that as a car moves along the road network, there are a large number of road features that can be identified based on their unique signature on different sensors. For example, Figure~\ref{fig:slm_ex1} shows the change of the phone orientation angle when the car moves along a curved road. The figure highlights that unique features (e.g. variance of orientation angle) can be extracted from the sensor trace to reflect unique anchors (e.g. transition from a straight to a curved segment) on the road. These unique anchors can be used to reset the phone location and hence reduce the error.

\emph{Dejavu} uses two types of anchors: physical anchors and virtual anchors.
Physical anchors can be mapped to road features and have uniquely identifiable sensor signatures based on a certain template. These include  bridges, tunnels, turns, curves, railway crossings, cat's eyes, speed humps, etc. Those anchors  locations can be extracted from the map or through prior knowledge. \emph{Dejavu} uses these physical anchors to seed its anchor database in a new city and bootstrap the error resetting process.

On the other hand, virtual anchors are detected automatically using unsupervised machine learning techniques to further extend the size of the anchor database, providing more error resetting opportunities and hence enhancing accuracy. These virtual anchors have a unique signature in the sensor signal space and does not necessarily have a physical association with a road feature. Examples include points with unique GSM or WiFi RSS signature, areas with anomalous sensor behavior, among others. Those virtual anchors and their locations are learned through a crowd-sourcing process in \emph{Dejavu}. 

\section{The Dejavu System}\label{sec:designdet}

In this section, we provide the details of \emph{Dejavu} covering sensors data preprocessing, physical anchor detection, virtual anchor detection, features selection, and anchor location estimation. We end the section with a discussion about different aspects of \emph{Dejavu}.

\subsection{Preprocessing}
Anchor detection in \emph{Dejavu} is based on consistent changes in the raw signal. To reduce the effect of noise and spurious changes, e.g. a sudden break or lane change, we apply a low-pass filter to the raw sensor data. In particular, we use local weighted regression to smooth the data \cite{cleveland1988locally}. 

\begin{figure}[!t]
\centering
\includegraphics[width=\figscale\linewidth]{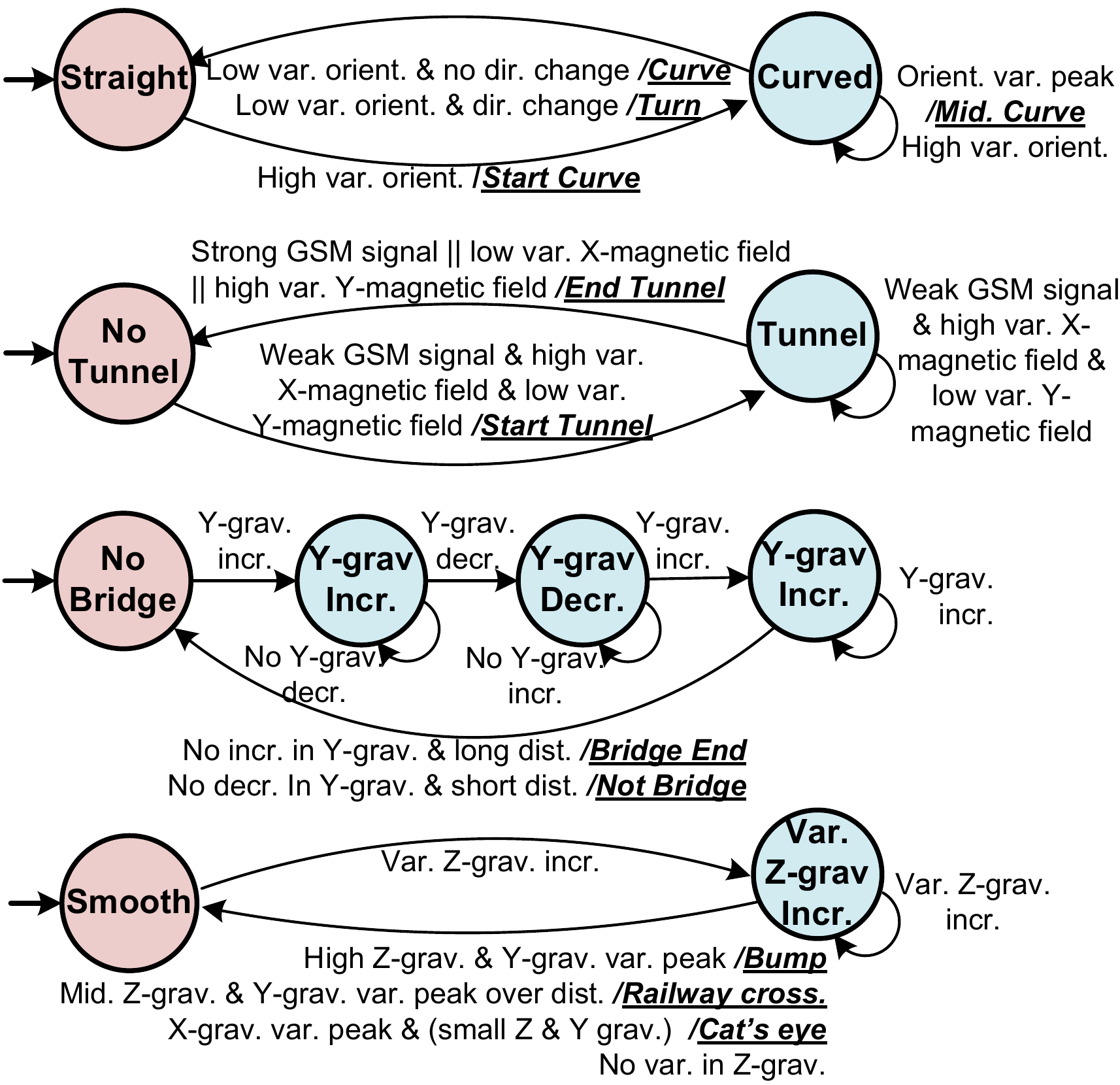}
\caption{A Mealy finite state machine for detecting different classes of physical anchors. The anchors are identified as output of the machine (highlighted in bold and underlined).}
\label{fig:slm_state}
\end{figure}
\subsection{Physical Anchors}
Physical anchors are used to seed the anchor database. These are anchors that can be identified from the map and/or prior knowledge of their locations. We could identify different classes of physical anchors that have a unique signature on the multi-modal raw sensor vector including bridges, tunnels, turns, curves, railway crossing, cat's eyes, speed humps, among others.  Note that fine-grained differentiation between different classes, e.g. separating curves from turns, is important as it reduces the confusion between adjacent anchors. Figure~\ref{fig:slm_state} gives a Mealy finite state machine that can be used to detect the different classes. In the balance of this section, we give details about identifying the different classes. Note that different sensors can be used to identify the same landmark, which is useful for increasing the detection accuracy.

\subsubsection{Curves and turns}
Road curvature, including curves and sharp turns, forces the car to change its direction, which results in a big variance in the phone's orientation (Figure~\ref{fig:slm_ex1}). This can be captured using the orientation angle.
To further differentiate between curves and turns we note that turns has a large difference between their start and end orientation angles. The associated anchors are detected at the beginning and end of the curve (when the variance goes above or below a threshold), at the point corresponding to the peak of variance, or at the location of the turn (angle change is above a threshold).

\subsubsection{Tunnels}
\begin{figure}[!t]
\centering
\includegraphics[width=\figscale\linewidth]{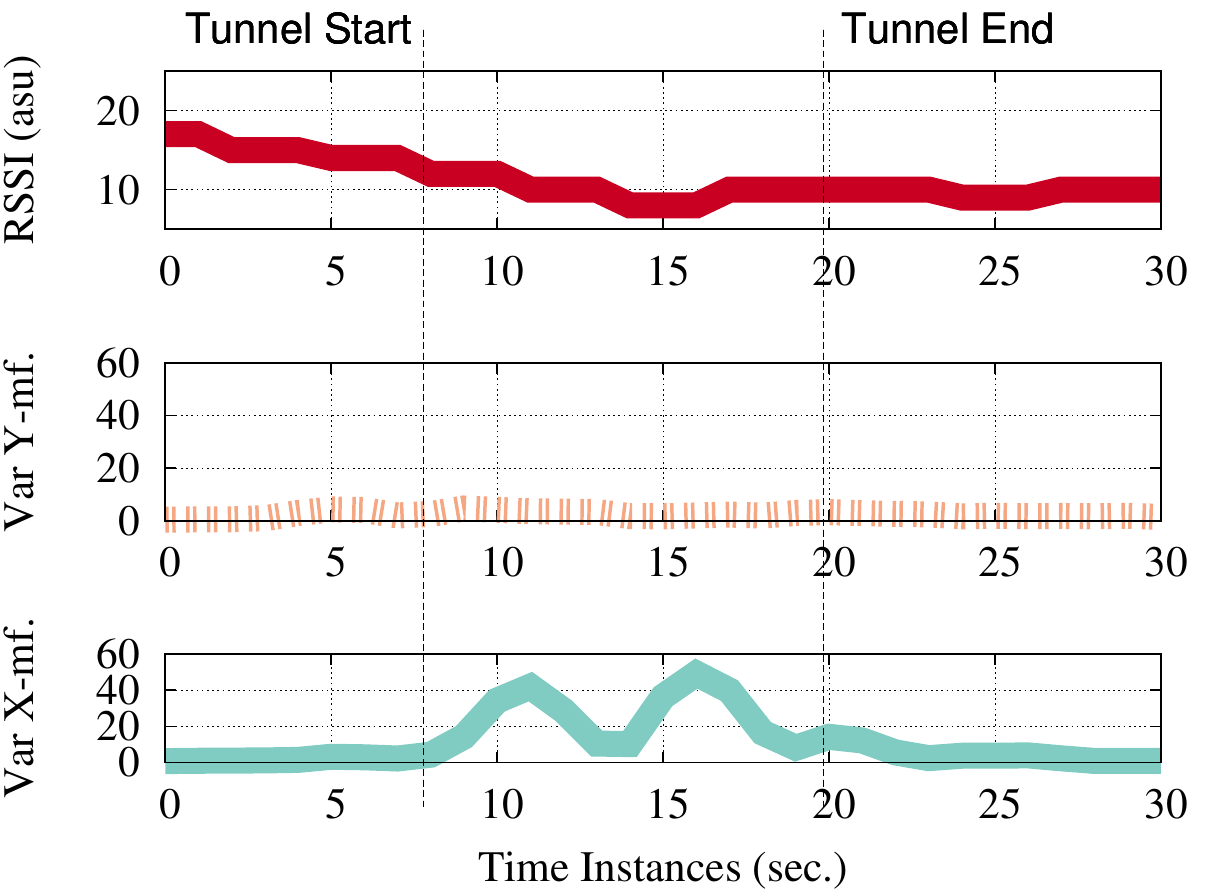}
\caption{Example of the different sensors behavior that characterize a tunnel: reduction in the cellular RSS, high variance of the ambient magnetic field in the x-direction, and low variance of the ambient magnetic field in the y-direction.}
\label{fig:tunnel}
\end{figure}

Going inside tunnels causes a drop in the cellular signals (for the associated and neighboring cells). We also noticed a large variance in the ambient magnetic field in the x-direction (perpendicular to the car direction of motion) while the car is inside the tunnel. This can be explained by the metal that exists on the side of the tunnel structure. Note that other classes of anchors may lead to a large variance in the magnetic field x-direction. 
  However, we found that tunnels are unique in having a large drop in the cellular RSS, high variance in the x-axis magnetic field, and low variance in the y-axis (direction of car motion) magnetic field as shown in Figure~\ref{fig:tunnel}. The associated anchors are detected at both the beginning and end of the tunnel.

\subsubsection{Bridges}
Bridges cause the car to go up at the start of the bridge and then go down at the end of the bridge. This is reflected on the y-gravity or z-gravity acceleration. Although other classes, such as bumps, cause the same effect (y or z gravity acceleration going up then down), bridges are unique in having this effect over a longer distance. The associated anchor is detected at the end of the bridge. Note that the bridge stretch marks are detected as another class of anchor.

\subsubsection{Road anomalies}
\begin{figure}[!t]
\centering
    \subfigure[X variance.]{
      \includegraphics[width=0.8\columnwidth]{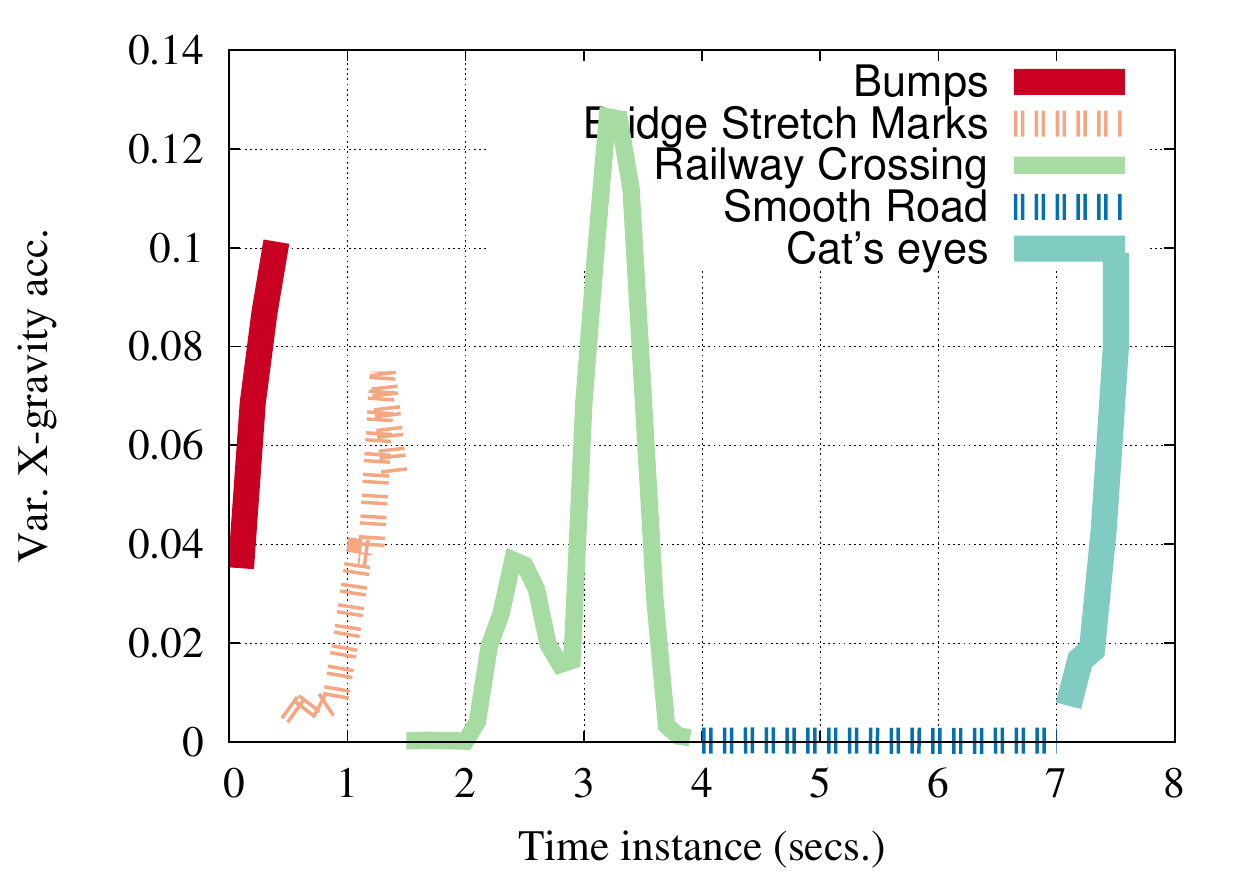}
      \label{fig:road_x}
    }
    \subfigure[Y variance]{
      \includegraphics[width=0.8\columnwidth]{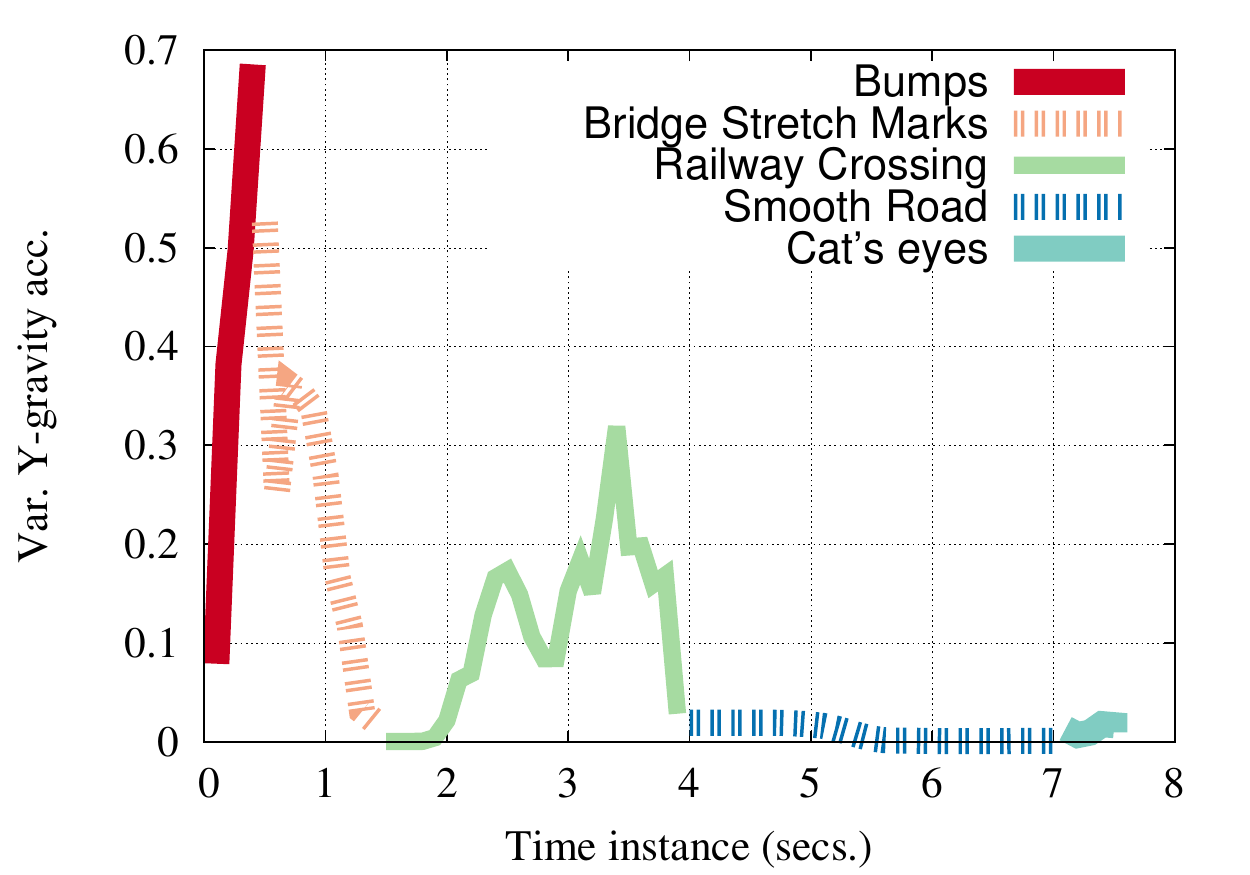}
  \label{fig:road_y}
    }
    \subfigure[Z variance]{
      \includegraphics[width=0.8\columnwidth]{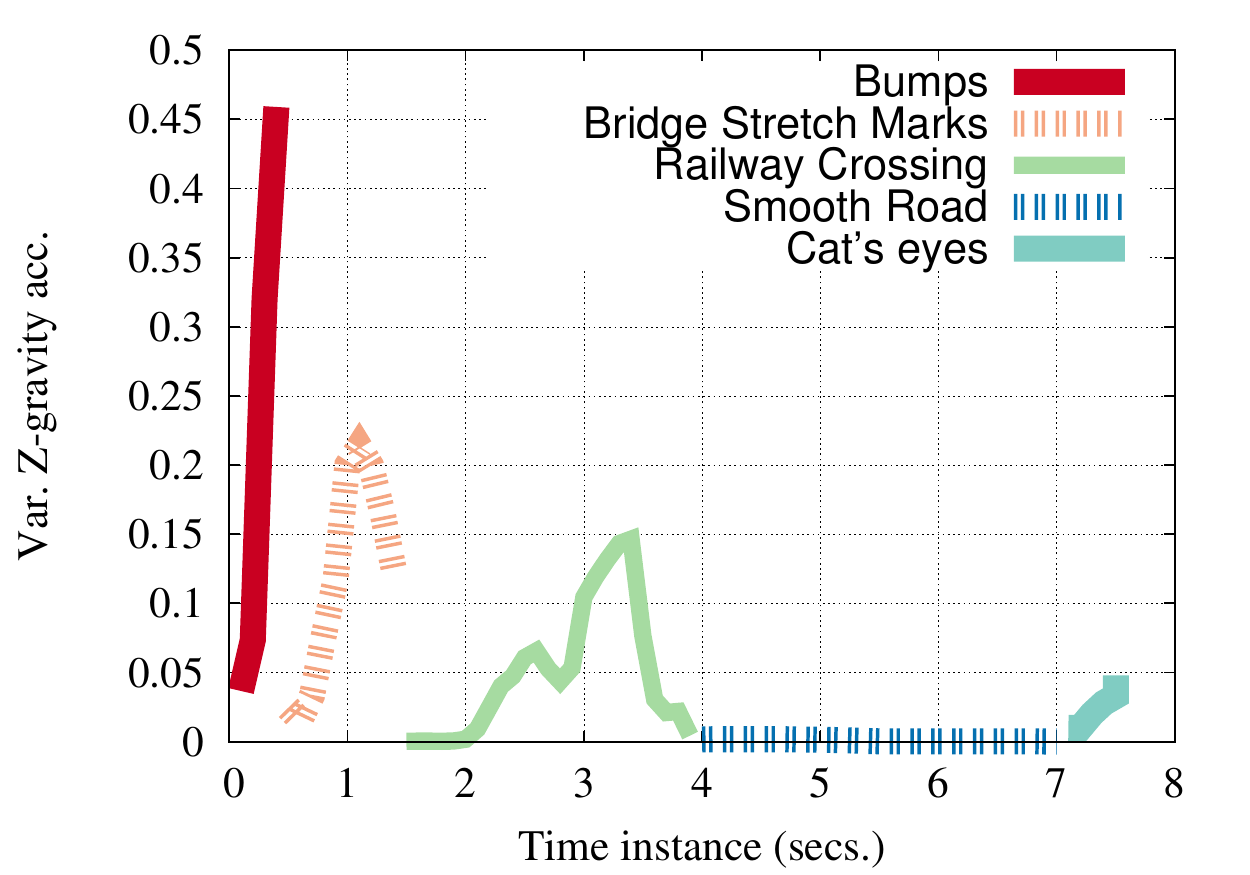}
  \label{fig:road_z}
    }

  \caption{Effect of different road conditions on the X, Y, and Z gravity acceleration variance. Cat's eyes have the lowest Y and Z variance, stretch marks are only detected inside a bridge, bumps have the highest Y and Z variances, while the railway crossing has a medium Y and Z variance. }
  \label{fig:road}
\end{figure}

Just like bridges, bumps, speed humps, cat's eyes, bridge stretch marks, and railway crossings all cause the car to move up then down, affecting all gravity acceleration axes. However, unlike bridges, all these classes affect the gravity acceleration over a small distance. To further separate these classes, we employ other sensors as follows (Figure~\ref{fig:road}):

{\bfseries Cat's eyes:} Unlike other road anomalies, the cat's eyes structure does not cause the car moving above them to have high variance in the y or z-axis gravity acceleration.

{\bfseries Speed humps:} usually have the highest variance in the y-axis and z-axis gravity acceleration compared to the other classes.

{\bfseries Bridge stretch marks:} These are only detected if the system detects that the car is on a bridge.

{\bfseries Railway crossings:} leads to a medium variance in the y-axis and z-axis gravity acceleration over a longer distance compared to other road anomalies.

\subsection{Virtual Anchors}
To further enhance accuracy and exploit other error opportunities, \emph{Dejavu} uses unsupervised learning techniques to identify virtual anchors that have distinct signature on cell-phone sensor readings along the road. These include points with unique GSM or WiFi RSS signature as well as anomalies in other sensors signatures.

Figure \ref{fig:arch} summarizes our virtual anchor detection technique.
First, anomaly detection techniques are used to identify anomalies in one or more sensors readings. Those anomalies are then clustered in the sensor space to identify candidate clusters of virtual anchors. Finally, the points of each cluster are then spatially clustered to identify the location of each anchor. An anchor is recognized if the cluster size is above a threshold and does not overlap with other instances of the same class within its surrounding area.
In the balance of this section, we start by discussing the selected features and then give the details of each submodule.

\subsubsection{Features selection}
\label{sec:features}
We have two main classes of sensors: vector-valued (e.g. WiFi and cellular) and scalar-valued sensors (e.g. x-gravity).

\vspace{2mm}
\noindent {\bfseries Vector-based sensors:}
\vspace{2mm}

Cellular towers information can be used as anchors. In particular, the GSM cellular specifications give the tower ID and the corresponding RSS for the cell tower the phone is associated with and up to six neighboring cells. This information is available through APIs in modern cell phones and presents an energy-free opportunity for obtaining ubiquitous virtual anchors.

Similarly,\emph{ when the WiFi interface is enabled}, users moving along the road hear WiFi access points (APs) from nearby buildings. 
Each WiFi sample corresponds to the list of APs MAC addresses and their corresponding RSS values.

Cellular and WiFi anchors correspond to points in the RSS signal space with unique signature.
To compute the distance between two samples in the cellular or WiFi feature space, we experimented with different fingerprinting techniques \cite{Horus,Radar}. We ended up using the following metric due to its low overhead and robustness to changes in the number of APs:
\begin{equation}
\frac{1}{\abs{A}} \sum_{\forall a \in A}^{} \frac{ min( f_1(a), f_2(a))}{max( f_1(a), f_2(a))}
\end{equation}
Where $A$ represents the union of the set of APs heard in the two samples; $f_i(a)$ represents the RSS heard from AP $a$ in sample $i$ ($f_i(a) = 0$, if $a$ is not heard in sample $i$). The similarity value of this metric is between 0 and 1.

The rationale for this equation is to add proportionally larger weights to the metric when an AP is heard at both locations with similar RSS. The normalization by the number of APs is used to make the metric more robust to changes in the number of APs between different locations.

\vspace{2mm}
\noindent {\bfseries Scalar-valued sensors:}
\vspace{2mm}

Each sample of these sensor streams can be represented by a single value, e.g. the gravity acceleration components. Note that even though some sensors are vector quantities, e.g. the gravity has three components along the $x$, $y$, and $z$ directions, we treat each of them individually to detect a richer set of anchors.

For a stream of a particular sensor readings, we extract features from a sliding window of size $w$, where features include the mean, maximum, variance, skewness, etc of the readings within the sliding window. More formally, denote the readings inside the sliding window starting at time $t$ as $S=(s_t,s_{t+1},...,s_{t+w-1})$, then each feature can be represented as $v=g(S)$, where $g$ is a function that maps the samples inside the window to a single feature value, e.g. the mean.

\subsubsection{Anomaly detection}
An anchor is defined as a unique point in its surrounding. To enhance the efficiency of anchor detection, we employ an anomaly detection technique on each of the scalar-valued sensors readings. In particular, we estimated the distribution ($\hat{f}(v)$) of the obtained feature sequence ($v_i$) as \cite{silverman1986density,bowman1997applied}:
\begin{equation}
\label{eq:kernel_density}
\hat{f}(v) = \frac{1}{nh}\sum_{i = 1}^{n}K(\frac{v-v_i}{h})
\end{equation}
where $h$ is the bandwidth, $n$ is the a sample size, and $K$ is the kernel function.
The choice of the kernel function is not significant for the results of the approximation \cite{scott2009multivariate}.  Therefore, we choose the Epanechnikov
kernel as it is bounded and efficient to integrate:
\begin{equation}
K(q)=  \begin{cases}
 \frac{3}{4} (1-q^2) , & \mbox{if } \left|q\right| \leq 1 \\
 0, & \mbox{otherwise}
\end{cases}
\end{equation}

 We also used Scott's rule to estimate the optimal bandwidth~\cite{scott2009multivariate}:
 \begin{equation}
   h^{*} = 2.345 \hat{\sigma} n^{-0.2}
 \end{equation}
where $\hat{\sigma}$ is the standard deviation of the feature stream.

After estimating the density function, we select critical bounds so
that if the feature values observed exceed
those bounds, the observed values are considered anomalous.
The critical bounds depend on the type of the feature selected;
Given a significance parameter $\alpha$ and assuming $\hat{F_j}$ is
the CDF of distribution shown in Equation~\ref{eq:kernel_density},
if the feature is a measure of central tendency (e.g. the mean), which can deviate
to the left or the right, then lower and upper bounds will be
calculated such that the lower bound is $\hat{F_j}^{-1}(\alpha/2)$
and the upper bound is $\hat{F_j}^{-1}(1 - \alpha/2)$ . However, if
the feature is a measure of dispersion (e.g. the variance), which can only deviate in
the positive (or right) direction, then an upper bound is only
needed and is equal to $\hat{F_j}^{-1}(1 - \alpha)$. We set $\alpha=0.4$ in our experiment as it balances the number of detected anomalies.

\subsubsection{Two stage clustering}
We cluster the features using hierarchical clustering in the vector feature space. This clustering stage will group similar anomalies, e.g. all potholes, into one cluster.

To identify the individual anchor instances within the same cluster, we apply a second stage of clustering; For a given cluster generated from stage one, we apply \textbf{\emph{spatial}} clustering to the points inside it based on the points coordinates.

To reduce outliers in both stages, a cluster is accepted only if the number of points inside it is above a threshold. In addition, a second stage spatial cluster is declared as an anchor if its points are confined to a small area and does not overlap with other clusters from the same anchor. The anchor location is taken as the weighted mean of the points inside the cluster.

\subsection{Computing the Anchor Location}
\label{sec:loc}

Whenever an anchor is detected, whether physical or virtual, its location is estimated as the current estimated phone location. However, since the user location has inherent error in it, this error is propagated to the anchor location.

To solve this issue and obtain an accurate anchor location, we leverage the central limit theorem. In particular, different cars will generally visit the same anchor through different independent paths. Therefore, averaging the reported locations for the same anchor from the different cars should converge to the actual anchor location.

\begin{figure}[!t]
\centering
\includegraphics[width=\figscale\linewidth]{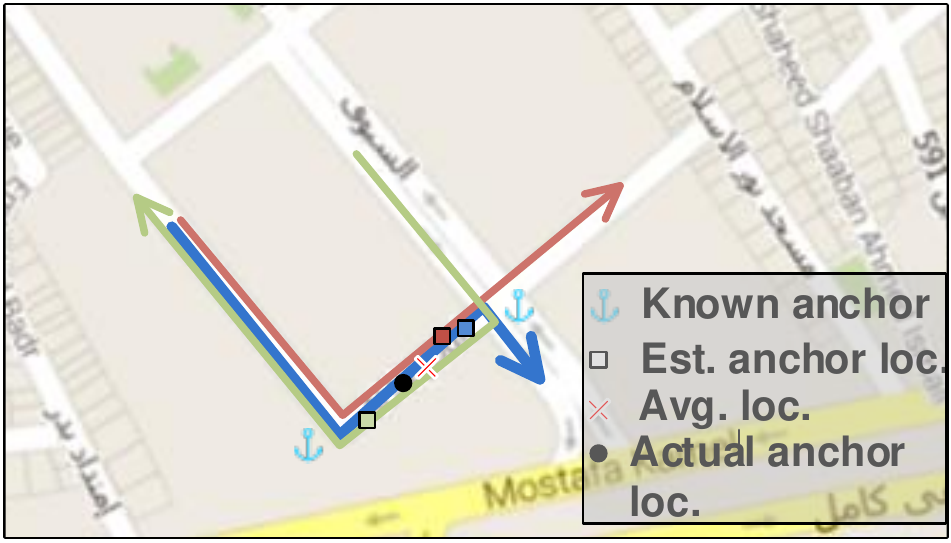}
\caption{Different car traces passing by the same anchor and the corresponding estimated anchor location. The starting point of each trace is the point of the last error resetting event. Shorter traces have higher accuracy.}
\label{fig:lm_loc_ex1}
\end{figure}

Moreover, we note that the longer the user trace before hitting an anchor from the last resetting point, the higher the error in the trace (Figure~\ref{fig:lm_loc_ex1}). Therefore, we use weighted averaging to estimate the anchor position, given  a higher  weight to shorter traces.

\subsection{Discussion}
\emph{Dejavu} provides accurate and energy-efficient outdoor localization suitable for car navigation. In this section, we discuss different aspects of the system. 
\subsubsection{Anchors aliasing}
Sometimes, different classes of anchors can be confused with other anchors at some samples. For example, a bump can be mistakenly detected as a railway crossing. To reduce this ambiguity, \emph{Dejavu} leverages the map context information. In particular, using the current estimated user location, one can filter out some anchor classes that do not exist in the map within a certain area around the user location, reducing ambiguity. We also note that some physical anchors can be detected as virtual anchors using the anomaly detection and the unsupervised learning engine. \emph{Dejavu} gives a higher priority to virtual anchors, which helps further reduce the ambiguity in the physical anchors detection, if any. 

\subsubsection{Efficient matching}
Similarly, we can leverage the user location, though not completely accurate, for efficient anchor matching. Particularly, \emph{Dejavu} limits its search space to a small area around the estimated user location. This significantly reduces the search space and increases the scalability of the system.

\subsubsection{Processing location}

\emph{Dejavu} can be split into a client-server architecture, where the sensors data is collected from the mobile and sent to the cloud for processing. Another model is to cache part of the anchors database  on the client, based on the current estimated location, and perform the matching locally on the client. Both techniques have their pros and cons in terms of required resources, latency, and communication cost and the optimal choice depends on the system designer's goals.

\subsubsection{Other sensors}
In this paper, we focused on the inertial sensors, cellular, and opportunistic WiFi due to their energy consumption advantage. Other sensors on the phone, such as the mic and camera, can be leveraged to further increase the density of anchors and hence accuracy. However, careful planning should be employed to study the energy-accuracy tradeoff. Sampling these sensors at a low duty cycle can be leveraged to reduce the energy consumption.

\subsubsection{Handling heterogeneity}
Different phones can have different sensor readings for the same anchor, especially in the range of readings. 
 To address this, \emph{Dejavu} implicitly applies a number of techniques including: anomaly detection based on the feature distribution, using scale-independent features (such as the variance), and combining a number of features for detecting the same anchor.

\begin{table*}[!t]
  \centering
    \begin{tabular}{||l||l|l|l||l||l|l||l|l|l||}
    \hline
    \hline
    \multicolumn{1}{||c||}{\multirow{3}[6]{*}{\textbf{Testbed}}} & \multicolumn{1}{c|}{\textbf{}} & \multicolumn{2}{c||}{\textbf{}} & \multicolumn{6}{c||}{\textbf{Average Anchor Density (per km)}} \\
\cline{5-10}    \multicolumn{1}{||c||}{} & \multicolumn{1}{c|}{\textbf{Distance}} & \multicolumn{2}{c||}{\textbf{Speed  (km/h)}} & \multicolumn{1}{c||}{\textbf{Phy.}} & \multicolumn{2}{c||}{\textbf{Virt. vect. sens.}} & \multicolumn{3}{c||}{\textbf{Virt. scalar sensors}}  \\
\cline{3-4}\cline{6-10}    \multicolumn{1}{||c||}{} & \multicolumn{1}{c|}{\textbf{covered  (km)}} & \multicolumn{1}{c|}{\textbf{Avg.}} & \multicolumn{1}{c||}{\textbf{Max.}} & \multicolumn{1}{c||}{\textbf{anch.}} & \textbf{GSM} & \textbf{WiFi} & \textbf{Acc.} & \textbf{Magnet.} & \textbf{Orient.} \\
    \hline
    \hline
    \textbf{City} & \multicolumn{1}{l|}{39.5} & \multicolumn{1}{l|}{12.6} & \multicolumn{1}{l||}{55.8} & \multicolumn{1}{l||}{3.3} & \multicolumn{1}{l|}{50} & \multicolumn{1}{l||}{112.2} & \multicolumn{1}{l|}{10} & \multicolumn{1}{l|}{7} & \multicolumn{1}{l||}{9} \\
    \hline
    \textbf{Highway} & \multicolumn{1}{l|}{50} & \multicolumn{1}{l|}{51.1} & \multicolumn{1}{l||}{100.1} & \multicolumn{1}{l||}{1} & \multicolumn{1}{l|}{33.3} & \multicolumn{1}{l||}{7} & \multicolumn{1}{l|}{2.9} & \multicolumn{1}{l|}{1.7} & \multicolumn{1}{l||}{1.4} \\
    \hline
    \hline
    \end{tabular}

  \caption{Summary of the different testbeds used. The high density of anchors, even without WiFi, allows Dejavu to obtain high-accuracy energy-efficient localization in both testbeds.}
  \label{tab:testbeds}%
\end{table*}%
\section{Evaluation}\label{sec:eval}
We implemented \emph{Dejavu} on different android devices including HTC Nexus One, Samsung Galaxy Note, Samsung Galaxy Nexus, and Samsung Galaxy S Plus. We evaluated the system in the city of Alexandria, Egypt as well as a number of major highways, covering a combined road length of 89.5~km. Table~\ref{tab:testbeds} summarizes the testbeds parameters.
Due to the low accuracy of the internal GPS for most of the used cell phones inside cities (as quantified in Section~\ref{sec:comp_loc}), we used an external bluetooth satellite navigation system that uses both the GPS and GLONASS systems as a \textbf{ground truth}.

For the rest of this section, we start by evaluating the accuracy of different system modules then compare the performance of \emph{Dejavu} in terms of accuracy and energy-consumption to the GPS and GAC~\cite{GAC} systems.
\subsection{Performance of the Different System Modules}
\subsubsection{Physical anchor detection accuracy}
Table~\ref{tab:slm_conf} shows the confusion matrix for detecting different physical anchors.
The table shows that different anchors
have small false positive and negative rates due to their unique signatures; The overall detection accuracy is 94.3\%.

\begin{table*}[!t]
\centering
\begin{tabular}{||c||c|c|c|c|c|c|c||c||c|c||c||} \hline
\hline
	& Cat's eyes	& Bumps	&Curves	&Rail cross. &Bridges	&Tunnels & Turns&unclass.&FP	&FN	&Total traces\\\hline\hline
Cat's eyes	& \cellcolor{gray!10}{\bfseries 22}&0&0&0&0&0&0&5&0&0.18	& 27\\\hline
Bumps	& 0&\cellcolor{gray!10}{\bfseries 30}&0&3 &0&	0&0&0&0.03 &	0.09 &	33\\\hline
Curves	&0&0&\cellcolor{gray!10}{\bfseries 20}	&0&0&0&0&0&0&0&20\\\hline
Rail cross. &	0&1	&0&\cellcolor{gray!10}{\bfseries 13}	& 0&0&0&0	&0.21&0.07 &	14\\\hline
Bridges	&0&0&0&0&\cellcolor{gray!10}{\bfseries 9	}&0&0&1&0&0.1	&10\\\hline
Tunnels	&0&0&0&0&0&\cellcolor{gray!10}{\bfseries 10}&0&0&0&0&10\\\hline
Turns	&0&0&0&0&0&0&\cellcolor{gray!10}{\bfseries 40}&0&0&0&40\\\hline\hline
Overall	&\multicolumn{8}{l}{}				 & \cellcolor{gray!20}{\bfseries 0.03}	& \cellcolor{gray!20}{\bfseries 0.06}	&\cellcolor{gray!20}{\bfseries 154}\\\hline
\hline\end{tabular}
\caption{Confusion matrix for classifying different physical anchors.}
\label{tab:slm_conf}
\end{table*}

\subsubsection{Virtual anchor detection accuracy}

\begin{figure}[!t]
\centering
    \subfigure[WiFi-City]{
      \includegraphics[width=0.45\columnwidth]{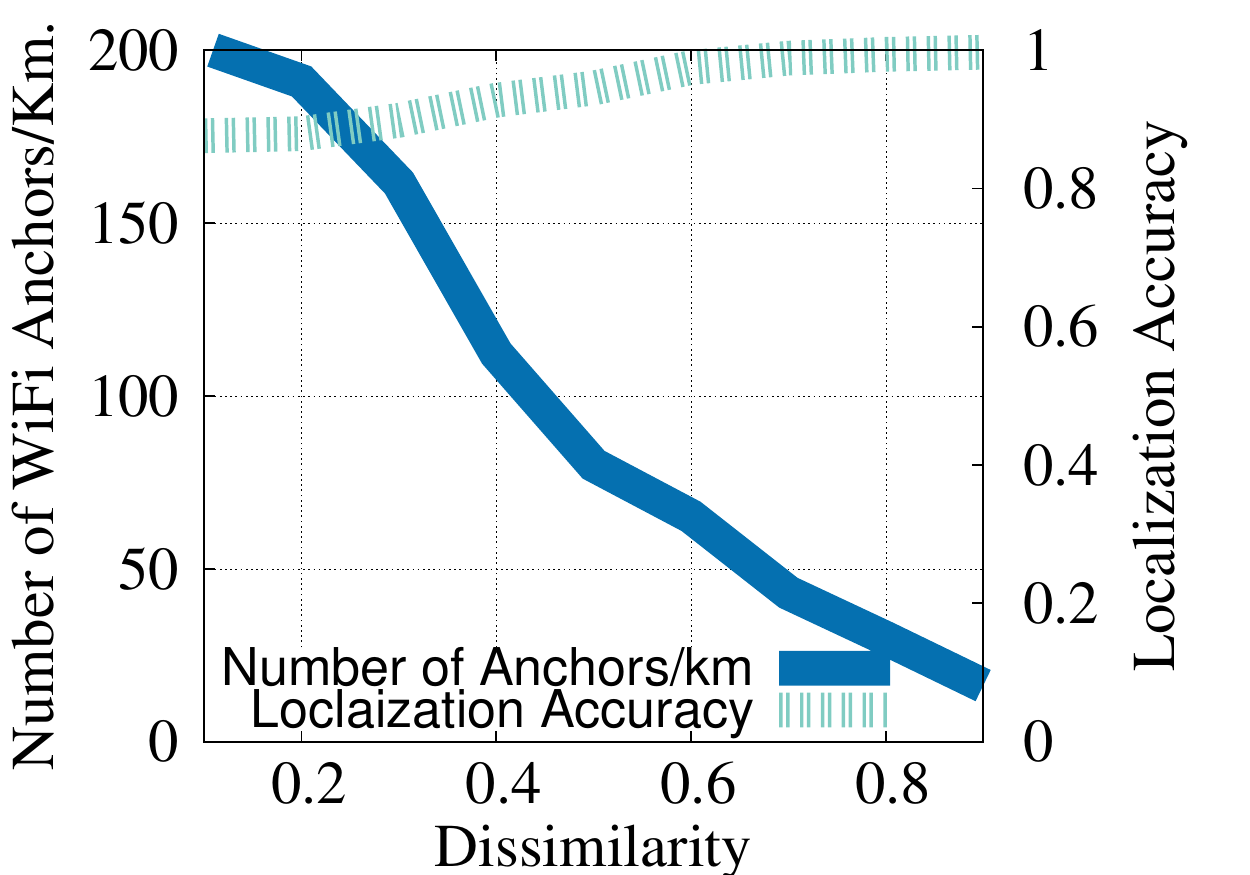}
    }
    \subfigure[WiFi-Highway]{
      \includegraphics[width=0.45\columnwidth]{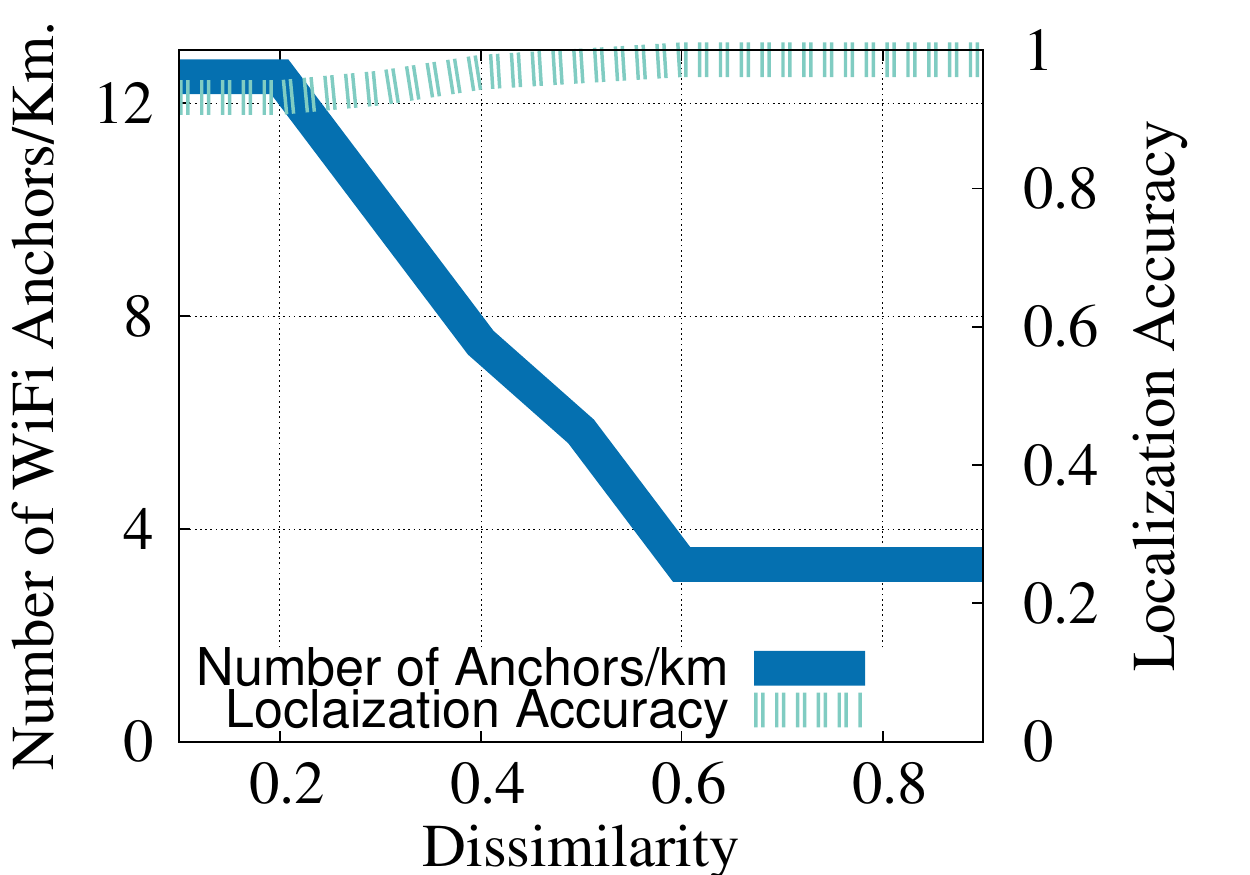}
    }
    \subfigure[Cellular-City]{
      \includegraphics[width=0.45\columnwidth]{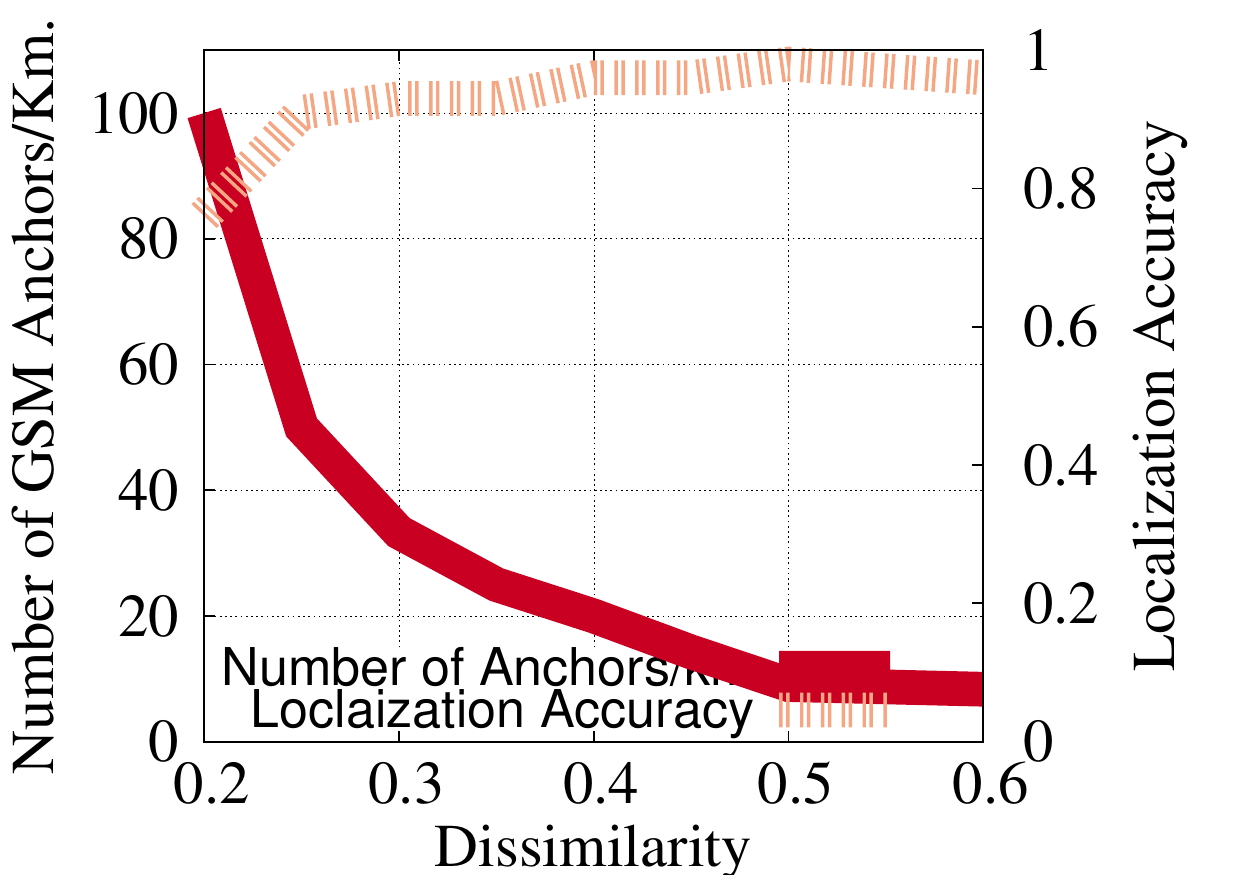}
    }
    \subfigure[Cellular-Highway]{
      \includegraphics[width=0.45\columnwidth]{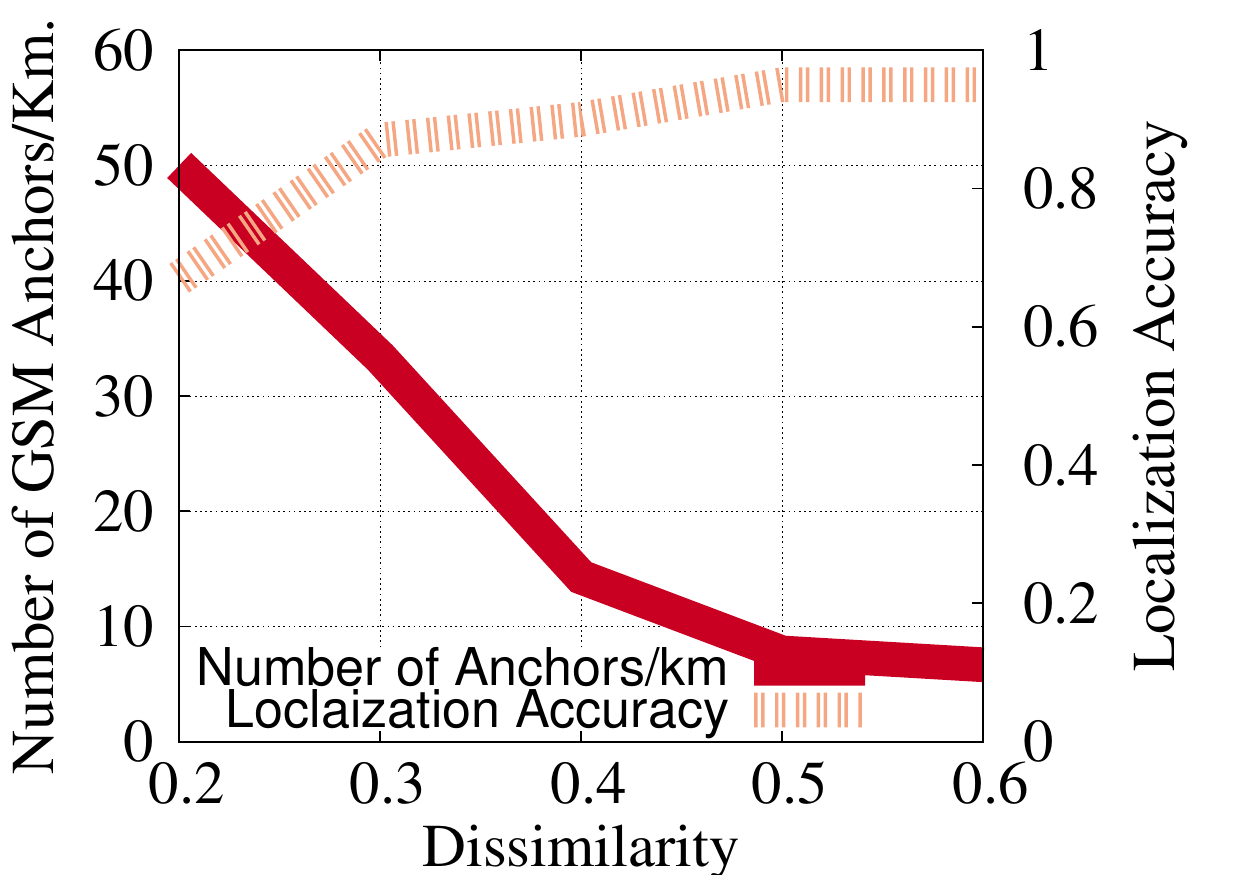}
    }
  \caption{Effect of changing similarity threshold on anchor density and correctly identifying the anchor}
  \label{fig:gsm_wifi}
\end{figure}

Figures~\ref{fig:gsm_wifi} shows the effect of changing the similarity threshold (Section~\ref{sec:features}) on the vector-based anchor density as well as the ability to differentiate between adjacent anchors.
This threshold is used in the first clustering stage to determine if two samples belong to the same cluster or not. Using a high value for the threshold leads to discovering more anchors.
The figure also shows that, due to the better signal propagation in highways, more distance is required to separate the anchors at the same similarity threshold.

 Comparing WiFi anchors to cellular anchors, WiFi anchors are more dense in cities due to their smaller coverage range and availability. However, since it is less probable to find WiFi APs on the highway, GSM anchors are more dense in this case.

We set the similarity threshold to (0.4, 0.4) for WiFi and (0.25, 0.3) for GSM for the (in-city, highway) cases. These values provide a good number of vector-based anchors while maintaining good differentiation accuracy.

\subsubsection{Anchor localization accuracy}
\begin{figure}[!t]
\centering
\includegraphics[width=\figscale\linewidth]{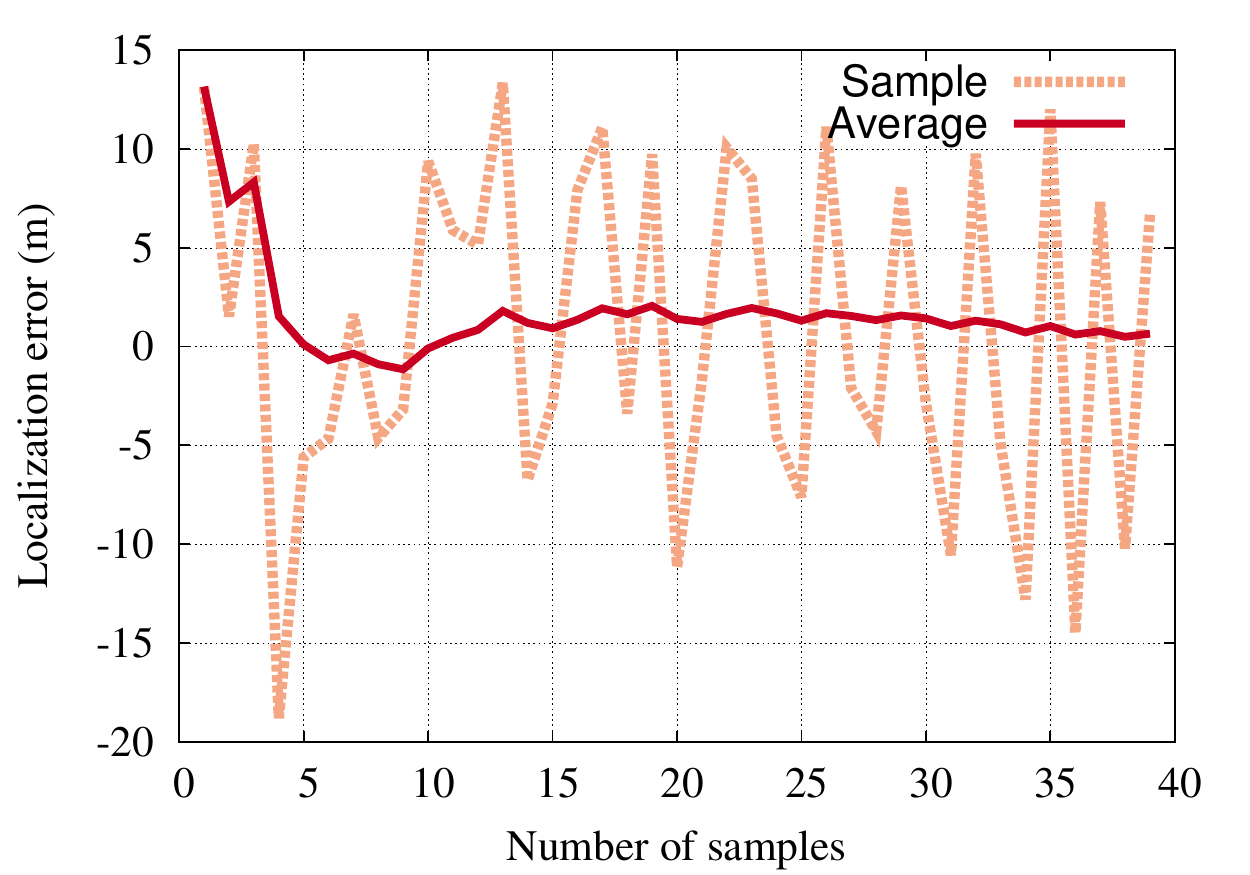}
\caption{Effect of number of samples on the accuracy of estimating the anchor location.}
\label{fig:users_wm}
\end{figure}

Figure~\ref{fig:users_wm} shows that the error in the anchor location drops quickly as we increase the number of samples. We can consistently reach an accuracy of less than 5m using as few as 20 samples for all discovered anchors. 

\subsubsection{Effect of anchor density on accuracy}
\begin{figure}[!t]
\centering
\includegraphics[width=\figscale\linewidth]{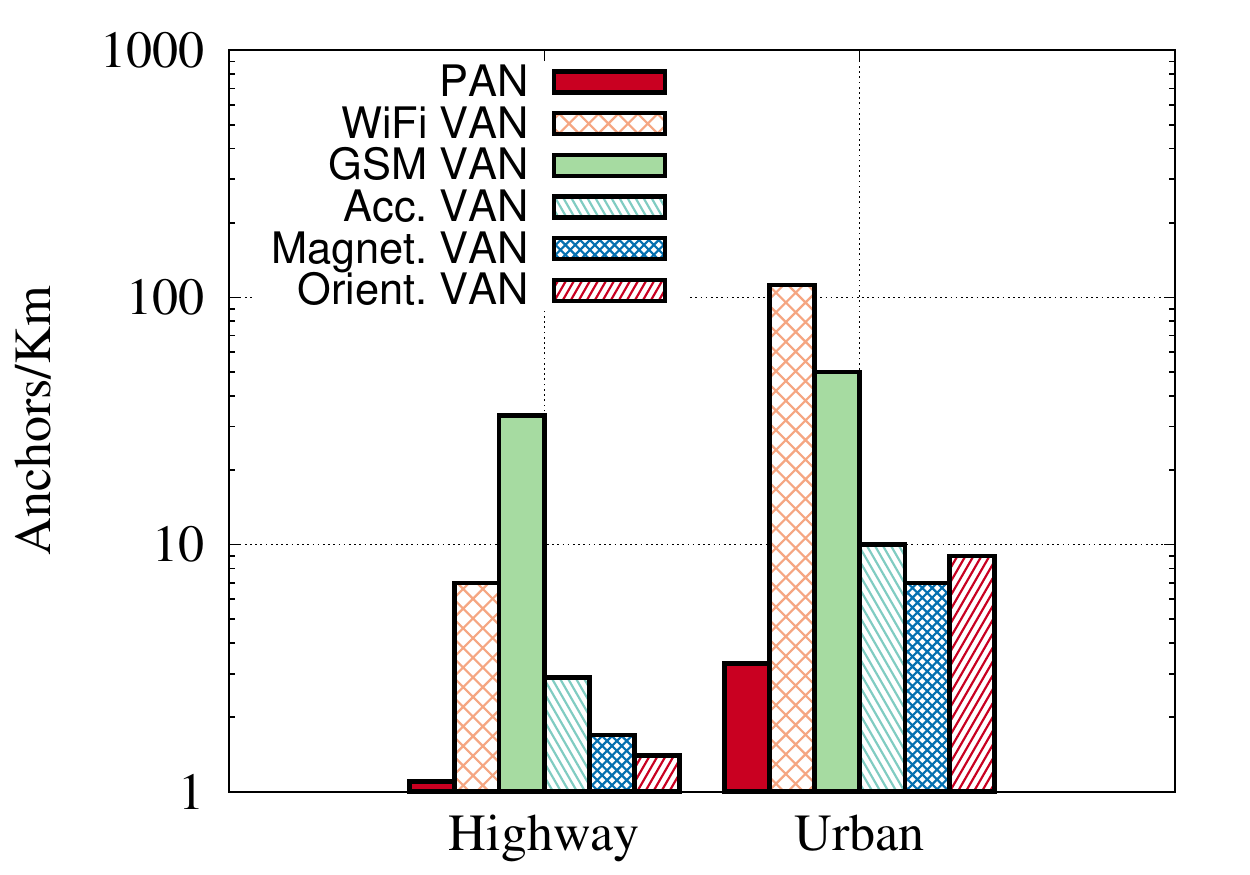}
\caption{Physical (PAN) and virtual (VAN) anchors density for the different classes in our testbeds.}
\label{fig:num_lms}
\end{figure}

Figure~\ref{fig:num_lms} shows the density of anchors in our testbeds. The figure highlights that there are indeed a large number of opportunities for error resetting, allowing \emph{Dejavu} to obtain accurate and energy-efficient localization. In addition, virtual anchors are much more dense than physical bootstrap anchors, highlighting their advantage.

\begin{figure}[!t]
\centering
    \subfigure[City]{
        \includegraphics[width=\figscale\linewidth]{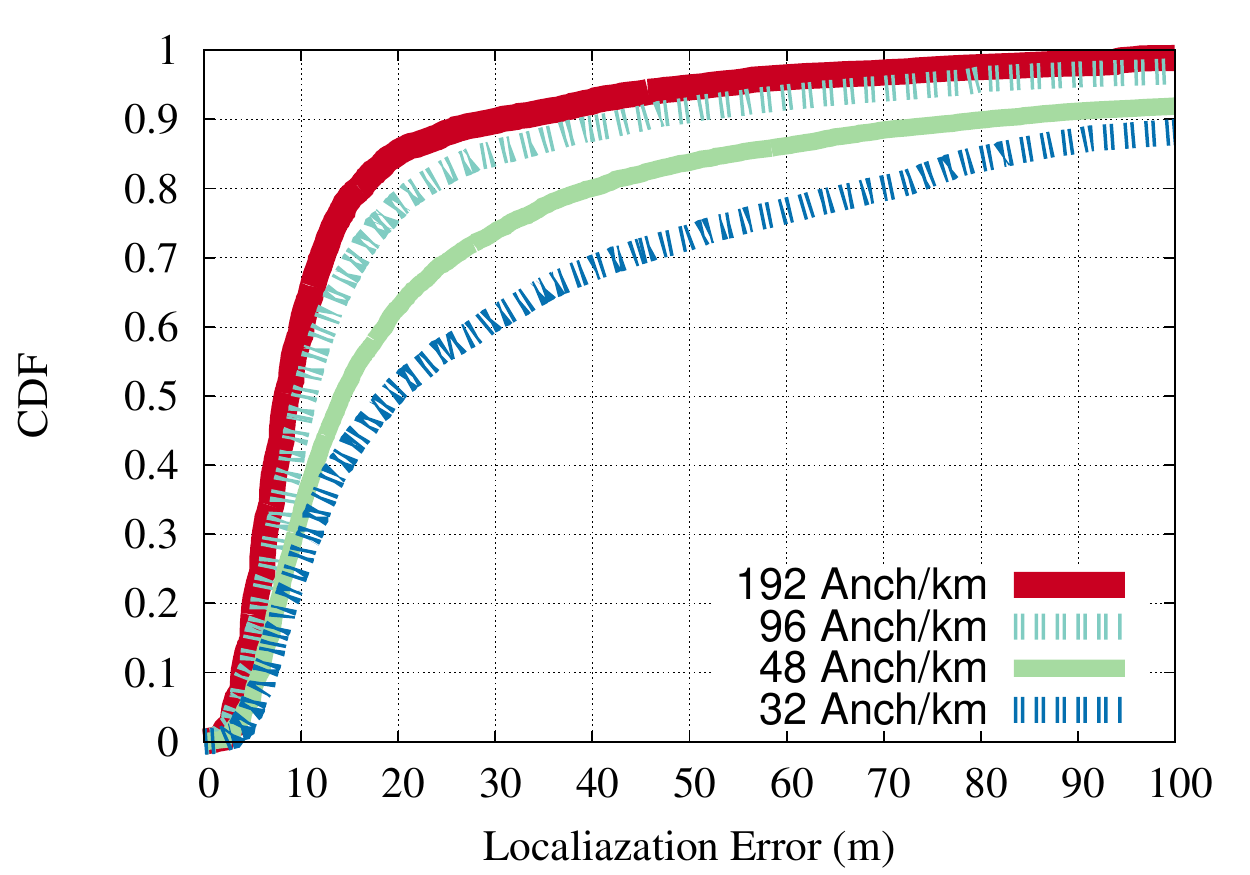}
    }
    \subfigure[Highway]{
      \includegraphics[width=\figscale\columnwidth]{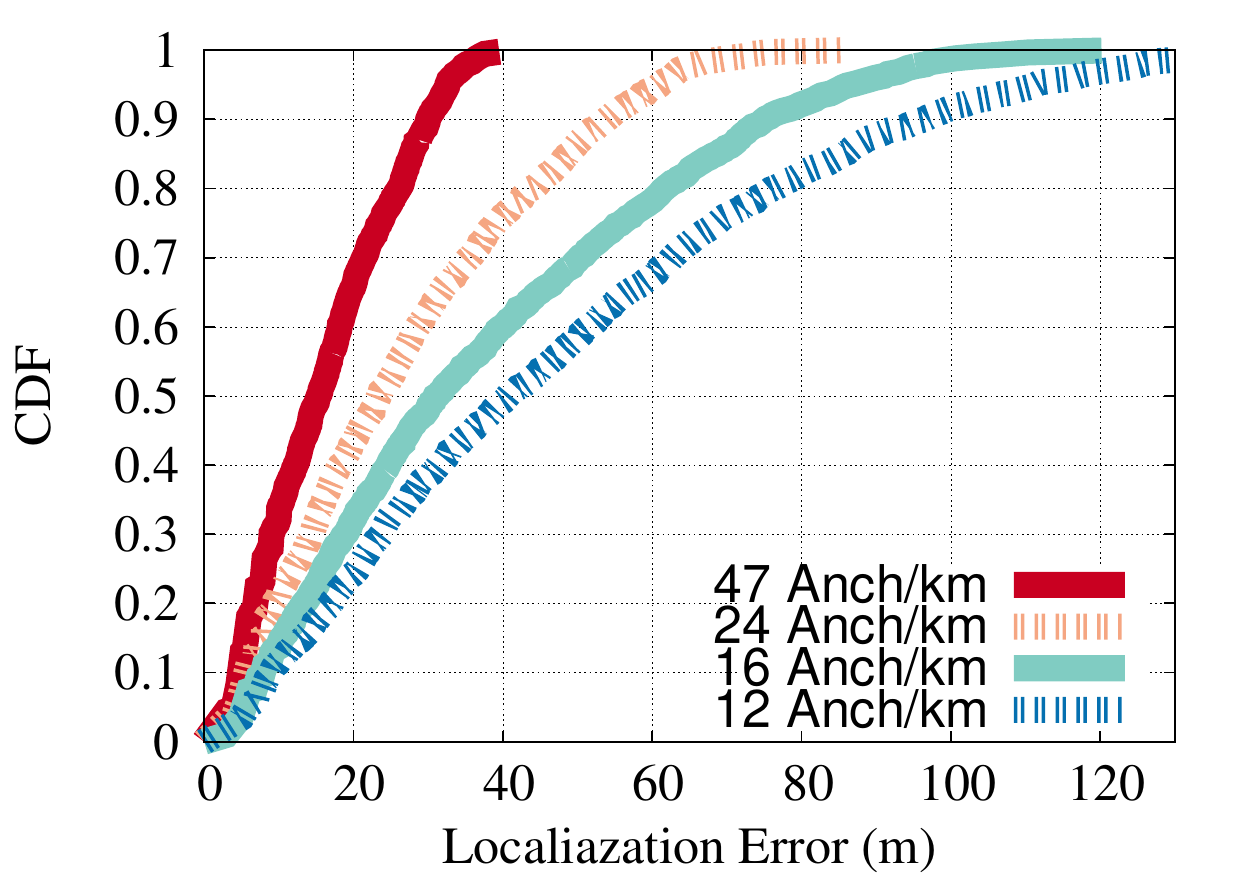}
    }
  \caption{Effect of density of anchors on the localization accuracy.}
  \label{fig:density_acc}
\end{figure}

Since we use WiFi opportunistically\footnote{We evaluate Dejavu performance without WiFi anchors in Section~\ref{sec:comp_loc}.} and the number of anchors detected from the sensors varies from one road to another depending on its characteristics,
we evaluate the effect of changing the anchor density on accuracy to generalize the obtained results to other areas. For this, we uniformly sample the detected anchors to obtain a specific density. Figure~\ref{fig:density_acc} shows the effect of anchor density on the system localization accuracy. We can see that even with a low density of anchors, as low as 28/km, the accuracy is still high (23m in city and 20m in highway).

\subsection{Comparison with Other Systems}
In this section, we compare the performance of \emph{Dejavu} in terms of accuracy and energy consumption to GPS and \emph{GAC}~\cite{GAC}. Similar to \emph{Dejavu}, \emph{GAC} uses dead-reckoning to estimate the phone location. However, to reset the accumulation of error, \emph{GAC} synchronizes with the GPS with a low duty cycle. Table~\ref{tab:comp}
 summarizes the results.

\subsubsection{Localization Error}
\label{sec:comp_loc}
\begin{figure}[!t]
\centering
    \subfigure[City]{
        \includegraphics[width=\figscale\linewidth]{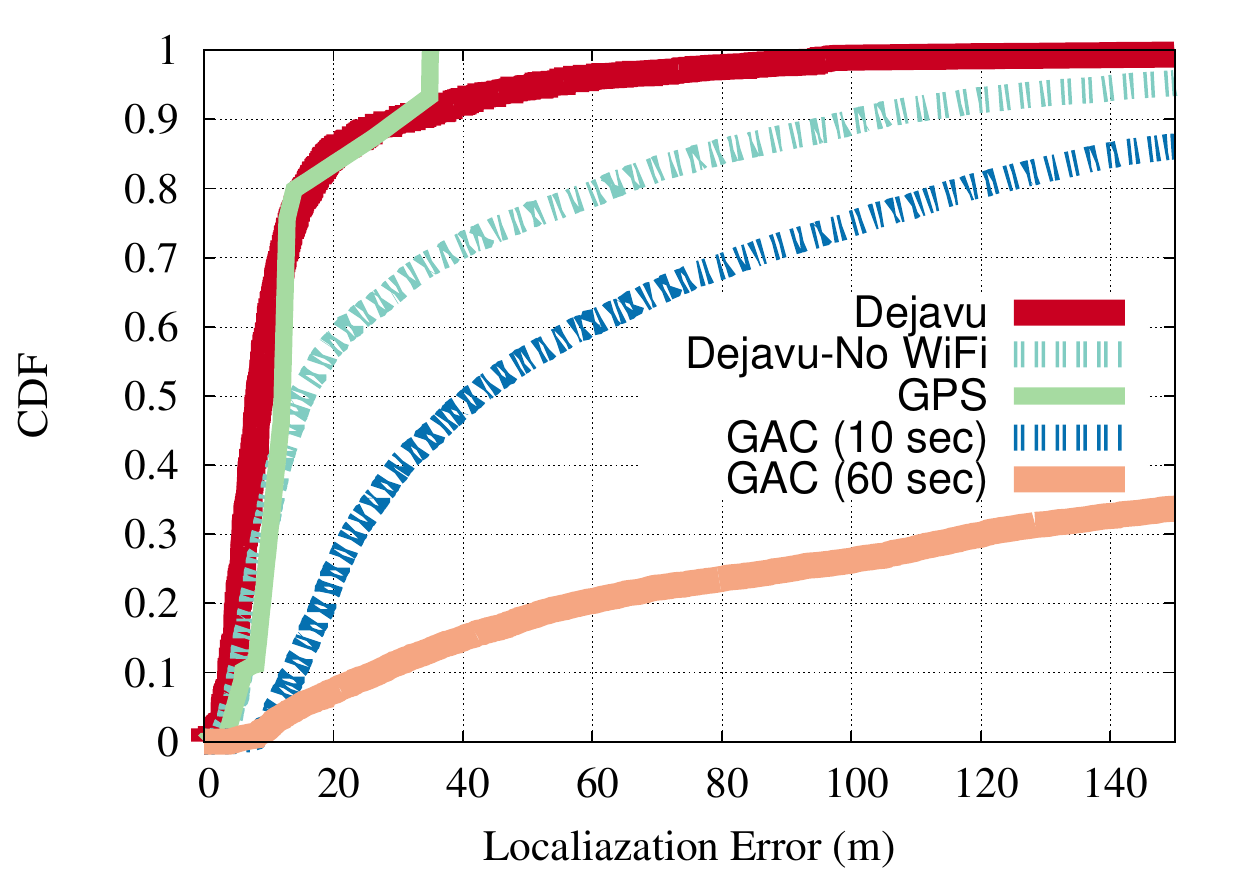}
    }
    \subfigure[Highway]{
      \includegraphics[width=\figscale\columnwidth]{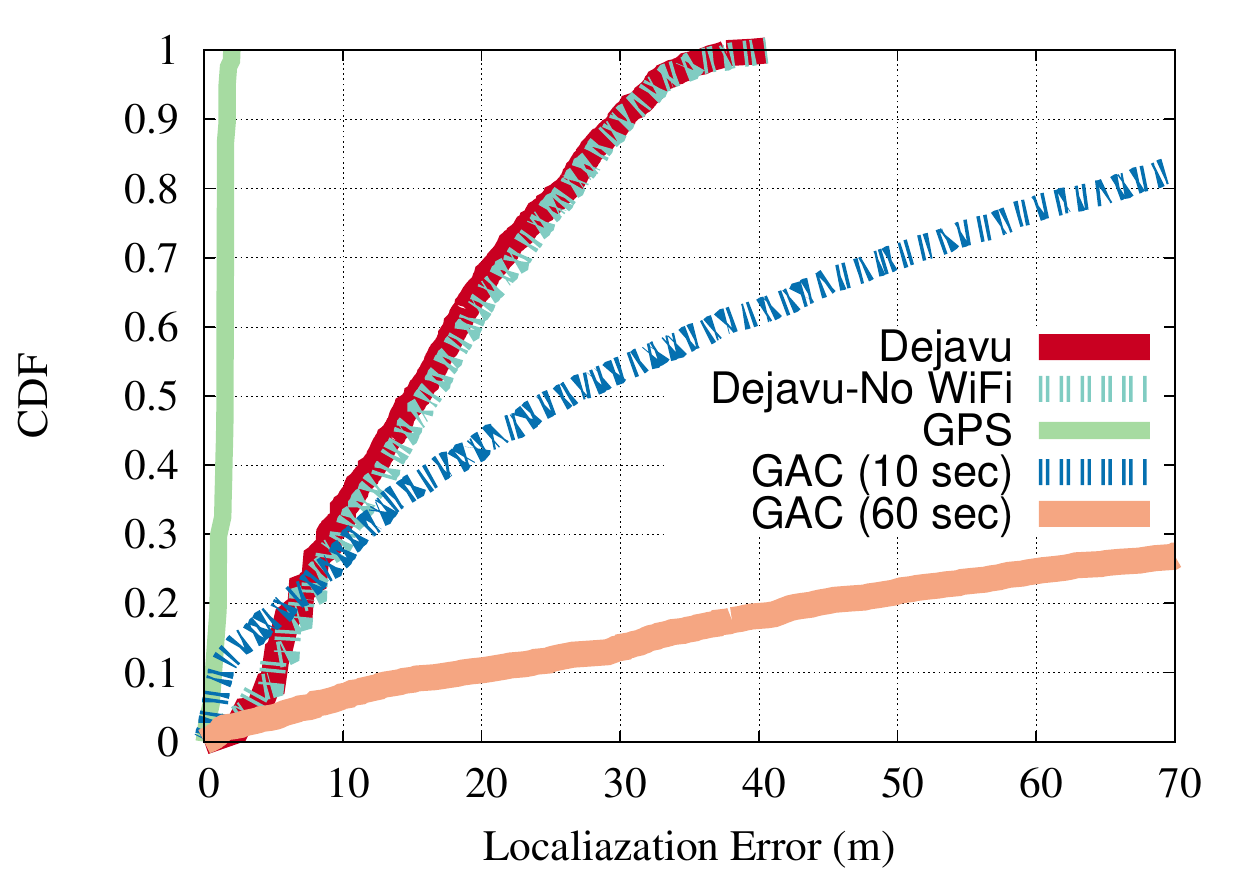}
    }
  \caption{CDF of localization error of \emph{Dejavu} compared to GPS and GAC~\cite{GAC}.}
  \label{fig:cdf_err}
\end{figure}

Figure~\ref{fig:cdf_err} shows the CDF of localization error for the three systems. The GPS accuracies are reported by the android API. The figure shows that \emph{Dejavu} gives better accuracy than GPS within cities, even without the WiFi anchors, due to the urban canyon problem and going inside tunnels or under bridges that affect the GPS accuracy. \emph{Dejavu} maintains its accuracy in highway driving, though its accuracy is slightly less than GPS. GPS, however, has the best worst-case error, as in rare cases, i.e. at the tail of the distribution, discovering anchors may be delayed, leading to accumulation of the error from dead-reckoning.

GAC~\cite{GAC} accuracy, on the other hand, depends on the duty cycle; ranging from good accuracy  (synchronization rate of once per 10 seconds) to hundreds of meters (synchronization rate of once every 60 seconds). This has a huge impact on the energy consumption though, as quantified in the next section. Note also that when GPS performs poorly, e.g. inside cities, GAC accuracy is also affected.

\begin{figure}[!t]
\centering
\includegraphics[width=\figscale\linewidth]{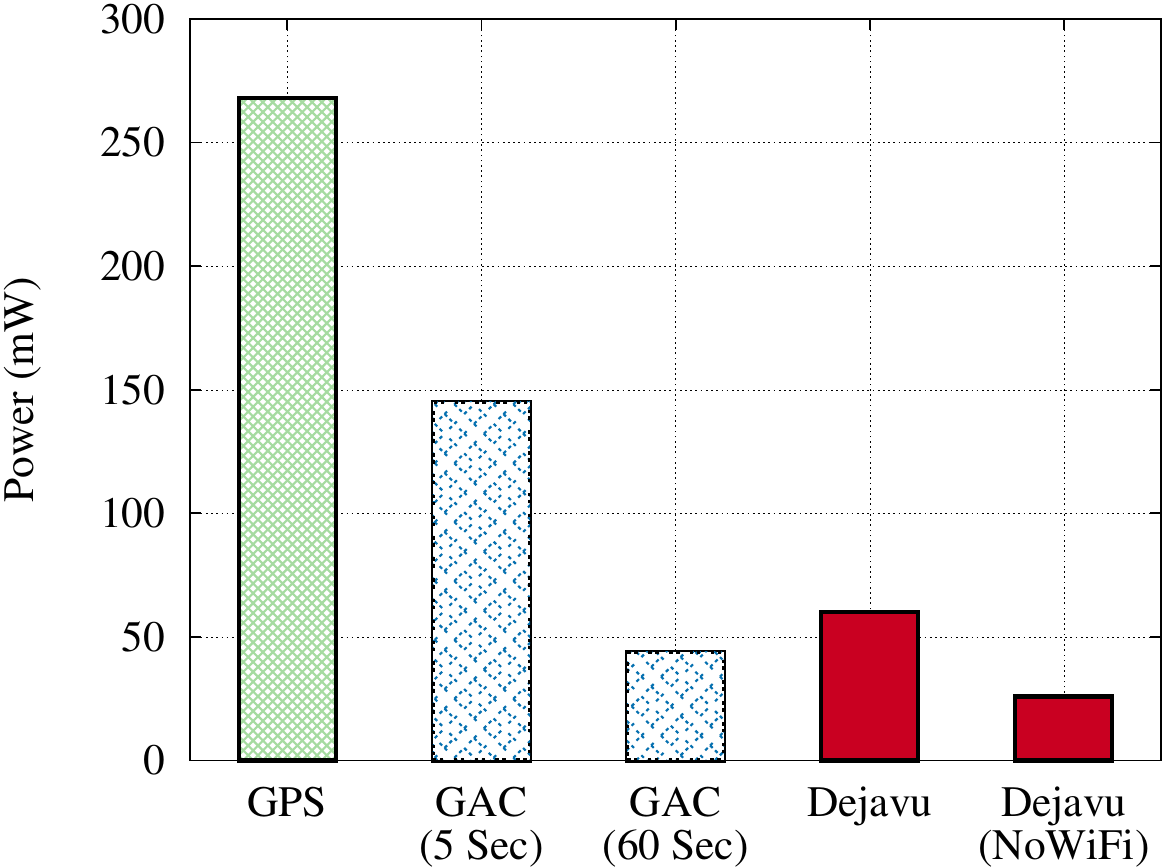}
\caption{Power consumption for the different systems.}
\label{fig:power}
\end{figure}

\subsubsection{Power consumption}
Figure~\ref{fig:power} shows the power consumption for the different systems calculated using the PowerTutor profiler \cite{zhang2010accurate} and the android APIs using the HTC Nexus One cell phone. The figure shows that \emph{Dejavu} has a significant advantage in power consumption, with 347\% saving in power compared to the GPS. The no-WiFi mode can further double this advantage.

\begin{table*}
\centering
\begin{tabular}{||c||l|l|l||l|l|l||l||}\hline \hline
\multirow{3}{*}{Technique} & \multicolumn{6}{c||}{Localization Error} & \multirow{3}{*}{Power Cons.$^*$} \\ \cline{2-7}
&  \multicolumn{3}{c||}{City} & \multicolumn{3}{c||}{Highway} & \\ \cline{2-7}
&Median & 75\% & Max. & Median & 75\% & Max. &\\ \hline \hline
Dejavu & 8.4 m & 13.8 m & 408.2 m & 16.6 m & 24.2 m & 41.5 m & 59.9  mW\\ \hline

Dejavu& 15.8 m & 48.5 m & 426.3  m & 18.2 m & 26.5  m & 41.9 m &25.9  mW \\
 (No WiFi) &(-88.1\%) & (-251.4\%) &	(-4.4\%) &	(-9.6\%) &(-9.5\%)	& (-0.96\%) & (+56.8\%)\\ \hline
 \hline
GAC & 41.7 m & 100.3 m & 538.3 m &  26.1 m & 57.3 m & 145.2 m & 109.7  mW \\

(10 sec) &(-390.6\%) & (-642.9\%) &	(-31.8\%) & (-69.5\%) & (-142.8\%)&(-257.6\%) & (-83.1\%) \\ \hline

GAC  &  302.8 m & 728.2 m & 2601.2 m & 168.8 m & 325.6  m & 822.9 m & 43.8 mW  \\

(60 sec) &(-3504.8\%) &	(-5176.8\%)	& (-537.2\%)&	(-916.9\%)	& (-1245.4\%) &	(-1882.9\%) & (+26.9\%)\\ \hline
\hline
\multirow{2}{*}{GPS} & 12 m & 12 m & 35 m &  1.3 m & 1.3 m & 2 m &  268 mW\\

 &(-42.9\%) &(+13.0\%) & (+91.4\%) & (+92.2\%) & (+94.6\%) &(+95.2\%) &(-347.4\%)\\ \hline
\hline\end{tabular}
\caption{Comparison between \emph{Dejavu}, GPS, and \emph{GAC}~\cite{GAC} with different duty cycles. Percentages are calculated relative to \emph{Dejavu}. $^*$~Power consumption is for the HTC Nexus One phone.}
\label{tab:comp}
\end{table*}

\section{Related Work}\label{sec:relwork}

\emph{Dejavu} combines dead-reckoning and sensor-based anchor detection to provide accurate and energy-efficient localization. Through this section, we discuss different techniques for outdoor localization and sensing.
The Global Position System (GPS)~\cite{hofmann1993global} gives an accurate location information in outdoor environments. However, it suffers from large power consumption and low accuracy when satellite signals are weak, e.g. due to tall buildings surrounding the road (urban canyons), inside tunnels, and bad weather conditions \cite{canyon1}. Even with sufficient satellites in view, the geometric dilution of precision can often be large with the estimated location points scattered around the original location~\cite{canyon2}, especially for the low-cost GPS chips typically available in cell-phones.

Alternative localization techniques have been introduced to address the GPS issues including city-wide GSM and WiFi localization that trade energy-efficiency with less accurate localization. 
 Typically these systems are cell-ID based~\cite{gmylocation} or use the RSS information \cite{cellsense2,cellsense,ibrahim2011hidden,wifiacc,skyhook,placelab,enloc,youssef2006location} as it is available for the application developers, compared to e.g. angle-of-arrival based systems. Due to the complex signal propagation effects, RSS-based techniques require the construction of a fingerprint map with RSS signatures for locations inside the area of interest. This calibration step is an expensive overhead, especially for covering large areas. Also, these systems suffer from low accuracy.

Recently, inertial sensors have become main-stream in cell-phones. 
 Researchers have leveraged these low-energy sensors to achieve energy-efficient localization \cite{unloc,uptime,CompAcc,GAC}.  The main idea is to use a dead-reckoning approach for localization and to reset its accumulating error. For example, by running the GPS sensor at a low-duty cycle. These systems still suffer from bad accuracy when the GPS signal is not available or when the GPS synchronization is more than 50 seconds, with errors as large as 100 meters inside cities and 250 meters in highway. \emph{Dejavu} eliminates the need of the energy-hungry GPS and leverages the array of independent sensors available on the phone for error resetting, achieving accuracies better than the GPS in some cases.

Map matching is a technique that has been used for error resetting in both research \cite{CompAcc,MapMatching} and commercial navigation products. It works by matching a GPS trace to the road network. However, this does not reset the displacement error unless there is a significant change in the trace, e.g. a turn. So if the car is moving in a straight line, the error will accumulate quickly with time. \emph{Dejavu} extends the map matching concept to matching with both physical and virtual anchors, which as we show in this paper are amble in both cities and highways, to achieve remarkable accuracy, even if the user is not changing direction and without using the GPS.

\emph{CrowdInside}~\cite{CrowdInside} is a system for the automatic estimation of indoor floorplans. Part of the \emph{CrowdInside} system is the generation of accurate indoor traces using error resetting with indoor anchors such as stairs and elevators. \emph{Dejavu} uses a similar approach for outdoor environments. However, outdoor anchors are completely different from indoor ones. In addition, we believe that the outdoor localization is a harder problem as the step pattern in the users' movement in indoor environments leads to less error accumulation compared to the outdoor car tracking. To compensate for this higher challenges, we introduce the virtual anchor concept to enrich the anchor database, allowing for higher error resetting opportunities.

Identifying road characteristics via different mobile sensors and monitoring road condition has been addressed in literature~\cite{pothole,nericell,mednis2011real}. These systems use the inertial sensors to monitor different road problems and use GPS to localize the sensed road problems. For example, the Pothole Patrol system~\cite{pothole} uses a 3-axis accelerometer and GPS to detect potholes along the road and separate them from  manholes and expansion joints. Similarly, Nericell~\cite{nericell} targets detecting road and traffic conditions using the phone sensors. They detect potholes using the same sensors. They did not differentiate between speed bumps and potholes as they assume that the location of intended speed bumps is known. Both Nericell and Pothole Patrol use external sensor chips which have higher sampling rates and lower noise compared to chips on typical cell-phones in the market. In~\cite{mednis2011real} authors presented an android based pothole detector which could identify different sizes of potholes.
\emph{Dejavu}, in comparison,  uses a larger set of sensors available on typical phones to detect different classes of anchors including tunnels, bridges, cat's eyes, among road anomalies. This is performed using a crowd-sourcing approach in a transparent way to the user. In addition, it employs an unsupervised learning approach in order to detect virtual anchors. Dejavu also uses standard sensors on typical phone with a low sampling rate for better energy-efficiency.

\section{Conclusion}
\label{sec:conclude}

We presented \emph{Dejavu}, a system for accurate energy-efficient outdoor localization suitable for car navigation. \emph{Dejavu} combines dead-reckoning with a novel anchor-based resetting technique to obtain accuracies that \emph{surpass GPS in city driving}. To achieve energy efficiency, \emph{Dejavu} depends on low-energy sensors and sensors that are already running for other purposes.
We showed how \emph{Dejavu} can extract and use both physical anchors from the map and virtual anchors using a crowd-sourcing based approach.

Implementation of \emph{Dejavu} on a number of android devices show that \emph{Dejavu} can achieve outdoor car localization with a median accuracy of 8.4m in city roads and 16.6m in highways. This is 42.9\% better in median localization error than GPS in city driving conditions, which is the most challenging case. In addition, \emph{Dejavu} consumes 347\% less power than GPS.

Currently, we are extending the system in multiple directions including outdoor pedestrian localization, extracting new anchor classes, using more phone sensors, among others.

\bibliographystyle{abbrv}
\bibliography{dejavu_arxiv_2}
\balancecolumns
\end{document}